\documentclass[useAMS,usenatbib,usegraphicx]{mn2e}
\usepackage{mathptmx}
\usepackage{amsmath}
\usepackage{verbatim}

\begin{document}

\title[Pulsar Radio Emission Region Exposed]{Radio Emission Region Exposed: Courtesy of the Double Pulsar}
\author[Lomiashvili and Lyutikov]{David Lomiashvili$^{1}$ and Maxim Lyutikov$^{1}$\thanks{E-mail:
lomiashvili@gmail.com; lyutikov@purdue.edu}\\
$^{1}$Department of Physics, Purdue University, West Lafayette, IN 47905, USA}
%\begin{document}

\date{}

%\pagerange{\pageref{firstpage}--\pageref{lastpage}} \pubyear{2013}

\maketitle

\label{firstpage}

\begin{abstract}
The double pulsar system PSR J0737-3039A/B offers exceptional possibilities for detailed probes of the structure of the pulsar magnetosphere, pulsar winds and relativistic reconnection.   We numerically model the distortions of the magnetosphere of pulsar B by the magnetized wind from pulsar A, including effects of magnetic reconnection and of the geodetic precession.  Geodetic precession leads to secular evolution of the geometric parameters and effectively allows a 3D view of the magnetosphere.
Using the two complimentary models of pulsar B's magnetosphere, adapted from the Earth's magnetosphere models by Tsyganenko (ideal pressure confinement) and Dungey (highly resistive limit), we determine the precise location and shape of the coherent radio emission generation region within pulsar B's magnetosphere.
We successfully reproduce orbital variations and secular evolution of the profile of B, as well as subpulse drift (due to reconnection between the magnetospheric and wind  magnetic fields), and determine the location and the shape of the emission region. The emission region is located at about 3750 stellar radii and has a horseshoe-like shape, which is centered on the polar magnetic field lines. The best fit angular parameters of the emission region indicate that  radio emission is generated on the field lines which, according to the theoretical models, originate close to the poles and carry the maximum current. We resolved all but one degeneracy in pulsar B's geometry. When considered together, the results of the two models converge and can explain why the modulation of B's radio emission at A's period is observed only within a certain orbital phase region. Our results imply that the wind of pulsar A has a striped structure only 1000 light cylinder radii away. We discuss the implications of these results for pulsar magnetospheric models, mechanisms of coherent radio emission generation, and reconnection rates in relativistic plasma.
\end{abstract}


\section{Introduction} \label{sec:Intro}

The first pulsar was discovered more than forty years ago \citep{hewish68}; however the exact mechanism of the pulsar radio emission is still uncertain \citep{mel95, mel10}. Moreover, not only the mechanism is uncertain but also the location of the emission generation is the subject of an active debate \citep{usov99, usov06, mel2000}.

Determining the  location of the emission region is one of the best probes for the pulsar radio emission mechanisms.
In terms of the radio emission height, there are two main classes of the models. First one places the emission height close to the neutron star surface ($\sim 10-100 R_{NS}$) \citep{gil03, musl04, wang12}, while the second one puts the emission at high altitudes ($> 1000 R_{NS}$) \citep{kaz87, lyut99a}, closer to the light cylinder. The main goal of this paper is to accurately determine the location of the radio emission region.

So far the estimates  of the emission heights have been done primarily  using radio polarization data combined with the rotating vector model \citep{Radha69} and the pulse profile widths \citep{gil93,kijak97,kijak03}. \citet{ganda01} and \citet{Dyks04a} have also proposed a phase-shift method to determine the emission height. In general, these methods show that core component emission originates very close to the surface of the neutron star (NS), but the conal components come from well above the surface \citep{rankin90,mitra02}. However, these techniques suffer from the significant uncertainties, mainly due to the ambiguities in the inferred pulsar geometries derived from the observational data. In addition, they are limited in that we observe only a small section of the magnetosphere of these isolated pulsars due to an unchanging line-of-sight (LOS). All this has changed after the discovery of the double pulsar.

The discovery of an eclipsing double pulsar system PSR J0737-3039A/B \citep{burg03, lyne04} has been hailed as a milestone in the field of astrophysics. The system consists of the fast recycled pulsar PSR J0737-3039A  (hereafter pulsar "A") with a period of $P_{A} = 22.7\, \mathrm{ms}$ and the slower but younger pulsar PSR J0737-3039B (hereafter pulsar "B") with a period of $P_{B} = 2.77\, \sec$, circling each other in the tightest known binary neutron star orbit of $2.4$ hours. This makes the double pulsar the best available test for general relativity and alternative theories of gravity \citep{kram06}.

The double pulsar system PSR J0737-3039A/B is very rich in observational phenomena, such as eclipses, orbital modulation of coherent radio and X-ray emission, and drifting subpulses \citep{kram08}. This provides a golden opportunity to verify and advance the models of pulsar magnetospheres, mechanisms of pulsar radio emission generation, and properties of the relativistic pulsar winds.

Along with the other outstanding features, what really sets the double pulsar apart is a simultaneous occurrence of such phenomena as eclipses of A by B, orbital modulation of B's radio emission, and  precession of B's spin axis. As we will show in this paper, all these features are independent from each other and have different origins. It is the simultaneous presence of these phenomena that makes it possible to determine the location and shape of the emission region.

It appears that our line of sight is nearly parallel to the orbital plane of the system, which leads to an eclipse of A by B's magnetosphere that is observed once per orbit \citep{lyne04, mcl04c}. Lyutikov and Thompson's (2005) \nocite{lyut05dp} and later Breton and colleagues' (2008) \nocite{breton08} detailed modeling of the eclipses allowed them to estimate pulsar B's geometry with an exceptional precision. Which significantly narrows down the parameter space of the model for B's magnetosphere.

Analysis of the observational data taken over several years revealed a pulse profile evolution in pulsar B. As \citet{per10} showed, presence of a precession along with a horse-shoe shaped emission beam can explain the observed modulation of the pulse profile widths. Estimated rate of geodetic precession agrees within $2\sigma$ confidence range with previously obtained value of $4.8(7)^{\circ} \mathrm{yr}^{-1}$, derived by \citet{breton08} from eclipse modeling. Furthermore, both estimates are consistent with the prediction of the general relativity $5.061(2)^{\circ} \mathrm{yr}^{-1}$. Presence of the precession gives us an invaluable information about the structure of the emission beam and the magnetosphere in general. Precession causes the change of the angle between the plane of the sky and B's spin axis. Therefore, we see different "cuts" of the magnetosphere by our line of sight,  after every revolution of the pulsar. This difference is extremely small and unobservable for two consecutive revolutions, however it is significant over the course of precession period, or even over a year \citep{per10}.

Unlike pulsar A, which maintains very steady emission (aside from the 30 sec duration eclipse), pulsar B exhibits
extreme variations in its flux density over a single orbit \citep{lyne04}. Bright single pulses are detectable and
can be studied in detail at two orbital phase regions, so-called bright phases. The bright phase 2 (hereafter BP2) appears near inferior conjunction (when pulsar B is between pulsar A and an observer, with a corresponding orbital phase of $270^{\circ}$) and ranges from $265^{\circ}$  to $305^{\circ}$, while the bright phase 1 (hereafter BP1) ranges from $185^{\circ}$ to $235^{\circ}$. In addition, pulse profiles have different shapes in the two orbital windows. \citet{lyut05bp} argues that the pulsar has the same intrinsic radio intensity throughout the orbit and that the orbital modulation is due to the deflection of the magnetic polar field lines with respect to the line-of-sight because of the influence of A.

In addition, fast intensity fluctuations similar to the drifting subpulses observed in B's pulsed emission provide the direct evidence of the influence of A's wind on B's magnetosphere \citep{mcl04b}. The drifting features have a separation of $\sim 23\, \mathrm{ms}$ within a given pulse, equal to the pulse period of A. Moreover, frequency of intensity fluctuations is equal to exactly the beat frequency between the periods of the two pulsars, 0.196 cycles/period. We interpret the drifting subpulses as a result of the distortions of the polar field lines caused by the reconnection between B's magnetic field and striped wind from A. As we will show in section \ref{sec:subpulsedrift}, such representation allows us to probe the structure of B's magnetosphere as well as the properties of pulsar wind on the scale much smaller than in the pulsar wind nebulae.

Depending on the location of the radio emission region and the line of sight (and hence on the orbital position) an observer will detect different radiation signatures of the distorted magnetosphere. Inversely, by studying the orbital modulation and using a model of the distorted magnetosphere we can deduce the location of the emission region. We used a novel approach to describe the distortions of B's magnetosphere induced by the wind of A.

Similar to the Sun, pulsar A produces a strong enough wind to interact with magnetic field of B and shape its magnetosphere. The nature of this interaction will vary depending on the properties of the wind. For a wind that is dominated by particle flux, formation of a parabolic bow shock,
similar to Earth's, is expected. In this hydrodynamic confinement model, the shape of the Earth's magnetosphere is mostly determined by the pressure balance between the supersonic solar wind and the Earth's nearly dipolar field. This parabolic shape of the magnetosphere is reproduced well by
current numerical models \citep{tsyg02a,tsyg02b,tsyg07}.

On the other hand, for a strongly magnetized wind, reconnection between the wind and the companion's magnetic field lines must be considered. This results in an open structure for the whole magnetosphere, similar to the one originally proposed by \citet{dungey61} for planetary magnetospheres. In the case of the double pulsar, it is not clear whether a hydrodynamic confinement model or a reconnection model is more applicable due to the unknown composition of A's wind. However, we are mostly interested in the overall magnetic structure of B's magnetosphere. For this purpose, it is sufficient to discuss magnetospheric structure in the most basic terms, relying on the models of planetary magnetospheres. We consider two extreme, though complimentary, models of the Earth's magnetosphere:  analytical, reconnection model of \citet{dungey61}, hereafter D61, and the semi-empirical, fully screened model of Tsyganenko \citep{tsyg02a,tsyg02b}, hereafter T02.

The plan of the paper is as follows. In section \ref{sec:Morphology} we review the main outstanding properties of PSR J0737-3039B's radio emission.  We present the advanced, numerical model for B's magnetosphere in section \ref{sec:ScreenedMag}. In section \ref{sec:Method} we present a method which we use to pinpoint the emission region using the detailed structure of the magnetosphere and the shape of the emission beam. We analyze the results of the simulations in the section \ref{sec:Fitting}. In section \ref{sec:results} we report the results  of the model. We review the implications of the results in section \ref{sec:Implications}. In section \ref{sec:subpulsedrift} we describe the simple Dungey-type model for B's distorted magnetosphere and constrain the emission height by reproducing the observed subpulse drift.
In section \ref{sec:Discussion} we discuss how our results for pulsar B test the radio emission mechanisms, as well as, list a few caveats of the model. Finally in section \ref{sec:Summary} we summarize the results and implications of our work.

\section{Morphology of PSR J0737-3039B's Radio Emission} \label{sec:Morphology}

The discovery of the double pulsar along with most of its observational properties was reported in two landmark papers by \citet{burg03}, who reported the discovery of a millisecond pulsar PSR J0737-3039A, and
\citet{lyne04}, who reported the discovery of the second normal pulsar PSR J0737-3039B.

Further studies have revealed a number of outstanding features of the individual pulsars, as well as features of the binary system as a whole. In fact, the most interesting properties of the second pulsar PSR J0737-3039B occur due to the influence of its companion. In this section we review a number of remarkable features of pulsar B that have been instrumental in our study of determining the geometry and structure of the radio emission region.

\begin{figure}
\centering
\includegraphics*[width=1.0\linewidth]{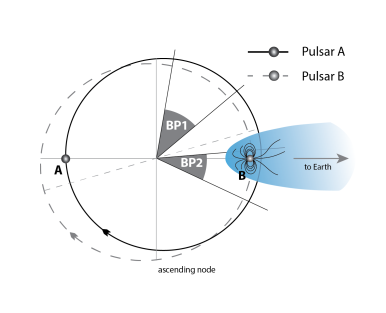}
\caption{Schematic view of the PSR J0737-3039A/B system orbit and the regions of brightened pulsed emission. Pulsar B displays a strong pulsed flux within BP1 and BP2 orbital phase intervals represented by the dark grey shaded slices. Pulsar A is visible along the entire orbit, except for $\sim 30 \sec$ at the superior conjunction where its obscured by the truncated magnetosphere of B. The locations of the bright phases are fixed with respect to the direction of the orbital motion (which was chosen arbitrarily) and the location of the ascending node.}
\label{fig:orbit}
\end{figure}

PSR J0737-3039B is the first radio pulsar to exhibit an orbital modulation of its radio emission \citep{lyne04, kram08}. B is observable along most of the orbit; however, it is exceptionally bright at two $30^{\circ}-40^{\circ}$ wide orbital phase regions. These narrow orbital windows are called bright phase 1 (hereafter BP1) and bright phase 2 (hereafter BP2) and are centered around the orbital longitudes $210^{\circ}$ and $285^{\circ}$, respectively, as measured from the ascending node (see Fig. \ref{fig:orbit}). A patchy radio observability is believed to be the main reason that discovery of B took longer than that of A.

Lyne et al. 2004 was the first to report about the bursting nature of B's radio emission. Since then, a number of extensive studies of B's light-curves have been conducted by \citet{mcl04b,mcl04c}, \citet{per10},\citet{per12}.

% Pulse profile evolution, within BPs

Detailed, more sensitive analyses revealed that in addition to the pulse intensity changing with orbital phase, the shape of the pulse changes as well. At the beginning of BP1, the pulse shows the trailing component dominating the leading one, which fades away by the end of the burst. The second bright phase shows the pulses with two components of more or less equal amplitude.

% Pulse profile evolution, long-term

Along with the short-term (on timescales of orbital period or less) variations in the pulse shape, data also show the secular changes in the observed pulse shape and intensity \citep{per10}. The radio pulses from B gradually transformed from a unimodal average profile in December 2003 ($\sim$ MJD 53000) to a broader two-component profile over the next few years. \citet{lyut05bp} predicted such an evolution of B's pulse profile based on the assumption that the geodetic precession of B's spin axis would cause our line of sight to cut through the different regions of the emission cone. Moreover, this pattern can be used to constrain the shape of the radio emission region (which we will discuss in more detail in Section \ref{sec:EmRegion}). The aforementioned assumptions are supported by the recent analysis of pulse profile evolution by \citet{per10}, who calculated the rate of separation of two pulse profile peaks to be ($\sim 2.6^{\circ}$ yr$^{-1}$) for both bright phase regions. As expected, this value of the separation rate is of the same order as the predicted geodetic precession rate, providing strong evidence for the precession-induced pulse profile evolution.

%time-evolving bright phases

Long-term changes in the properties of the bright phases are not limited to the pulse profile evolution. Observational data also revealed the changes in their duration and orbital longitude \citep{burg05,per10}. Over the course of 50 months, the center of BP1 remained more or less constant around $210^{\circ}$, whereas its width shrunk from about $40^{\circ}$ to about $30^{\circ}$ (widths are calculated at $10\%$ of maximum intensity). Unlike BP1, BP2 gradually shifted in both measures; its center drifted from orbital longitude of $\sim 280^{\circ}$ to $\sim 293^{\circ}$ and its width shrunk from about $30^{\circ}$ to about $20^{\circ}$ by MJD 54550. These patterns, shown on Fig. \ref{fig:BPResults}, are frequency independent over the range between 680 MHz and 3030 MHz \citep{lyne04, pos04}.

Observed bright phases evolve with rates of the same order as the pulse profile evolution and geodetic precession rates. This leads to the suggestion that all secular variations in B's radio emission are due to the geodetic precession.

%disappearance
The lightcurves of the two bright phases also differ in intensity. $820\, \mathrm{MHz}$ observations on MJD 52997 resulted in mean flux densities of 0.95 and 0.65 mJy for BP1 and BP2, respectively. Repeated measurements in the consequent epoches revealed gradually decreasing flux densities for both bright phases. Furthermore, the radio emission in both bright phase regions vanished over time, albeit at different rates. Closely after MJD 54550 (March 2008), the mean flux densities for BP1 and BP2 reached zero \citep{per10}. Similar to the pulse profile evolution, this phenomenon can be attributed to the changing impact factor between the emission direction and the line of sight, which changes due to the precessing spin axis of pulsar B. However, the analysis explaining the disappearance of B will differ depending on whether we use a pure or strongly distorted dipole as a model of B's magnetosphere (see Section \ref{sec:cutoff}).

% subpulse drift

Another outstanding property of B's radio emission, found by \citet{mcl04b}, is its modulation at the timescale coinciding with the spin period of A. This modulation reveals itself in the narrow features drifting through the pulse window, similar to the drifting subpulses.

The possibility of A's beamed emission being a cause of this modulation was quickly discarded due to its two-pole nature \citep{ferd2013} corresponding to $88\, \mathrm{Hz}$ instead of $44\, \mathrm{Hz}$ observed for the drifting subpulses \citep{mcl04b}. This peculiar relationship between the characteristic timescales of the two pulsars is direct evidence of the influence of A on B. \citet{mcl04b} proposed that it is A's electromagnetic radiation that affects B's magnetosphere, producing the modulation of radio emission. They argue that the subpulse-like features are caused by the electromagnetic field itself rather than its intensity or pressure, which would result in the $88\, \mathrm{Hz}$ periodicity.

Whether the reason for this behavior is pure geometric or whether it involves variations in the emission mechanism remains unclear. The situation is further complicated by the fact that this modulation is only observable in BP1 and is completely absent in BP2 (at times when pulsar B was still detectable).

\citet{mcl04b} argue that the absence of drifting subpulses in the BP2 has to do with the orientation of B's parabolic magnetosphere, formed due to the A's wind interacting with B's magnetosphere (see Fig. \ref{fig:orbit}). Near the superior conjunction, B's radio pulses that are emitted towards the Earth propagate through the magnetotail and remain relatively undisturbed. On the other hand, as suggested by \citet{mcl04b}, the radio pulses emitted in BP1 travel through the magnetosheath filled with hot dense plasma and get modulated. The authors do not discuss the exact process of the modulation. However, it is apparent that geometric properties of the double pulsar system play an important role.

The drifting subpulses provide an invaluable opportunity to study the properties of the pulsar wind and its interaction with a companion's magnetosphere. In Section \ref{sec:subpulsedrift}, we present the model that elaborates on the ideas suggested by \citet{mcl04b} and explains the origin and features of the drifting subpulses.

%All the features of pulsar B discussed in this section have one thing in common:

One main purpose of this work is to explain the features of B's radio emission described above. These features have one thing in common: they strongly depend on the structure of B's magnetosphere and its distortions. Therefore, in order to understand the peculiarities of B's radio emission, we need a realistic model of a binary pulsar magnetosphere and its interaction with a companion's wind.

\begin{figure}
\centering
\includegraphics*[width=1.0\linewidth]{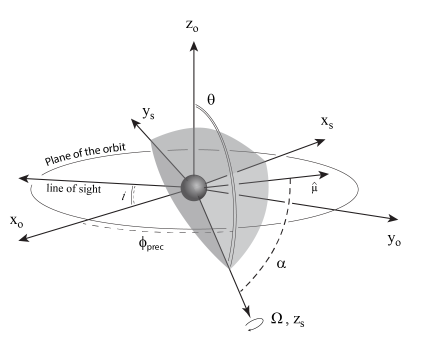}
\caption{Geometry of the pulsar B. The orbit of the PSR J0737-3039A/B is in the $x_{0} - y_{0}$ plane and its center coincides with the origin of the $(x_{0}, y_{0}, z_{0})$ cartesian coordinate system. The $x_{0}$ of the system always remains in the $x_{0}-\mathrm{LOS}$ plane. The orbital angular momentum is antiparallel to the $z_{0}$ axis. The spin axis $\Omega$ of pulsar B is inclined at angle $\theta$ to the orbital normal $z_{0}$ and at angle $\phi_{prec}$ with respect to the $x_{0} - z_{0}$ plane. $\alpha$ is a magnetic colatitude (i.e., the angle between the magnetic and spin axes). $i$ is the orbital inclination with respect to the plane of sky. Second coordinate system $(x_{s}, y_{s}, z_{s})$ is fixed with the dipole. Its $z_{s}$ axis is aligned with the spin axis, while the dipole axis $\mu$ lies in the $x_{s} - z_{s}$ plane.}
\label{fig:pulsargeometry}
\end{figure}

\section{Model of B's Screened Magnetosphere} \label{sec:ScreenedMag}

In addition to the usual suspects of pulsar physics such as: emission mechanism, location and shape of the emission region, another unknown factor in case of pulsar B is the structure of its presumably distorted magnetosphere.
Observational properties of pulsars A and B indicate that spin-down energy losses of A is about 3000 times greater than that of B. As a result, energetic relativistic wind from A blows away over $90\%$ of B's magnetosphere forming a nearly paraboloidal magnetopause (a boundary layer of shocked A's wind around B's magnetosphere) around pulsar B. Such cometary configuration would excite additional currents within and around the magnetosphere which in turn would induce the distortions of field lines close to the boundary. Resulted truncated magnetosphere extends from the pulsar B towards pulsar A ("dayside") at $\sim 0.3 R_{LC}$, while extends much further on the opposite - "nightside".

On the other hand, detailed modeling of A's eclipses by B confirmed that the structure of pulsar magnetosphere at intermediate distances ($R_{NS} \leq r \leq 1000 R_{NS}$) can be well represented by the simple dipole \citep{lyut05dp}. Therefore, pulsar B's magnetosphere is expected to retain a dipolar structure closer to the neutron star.

These predictions are supported by the advanced numerical simulations of magnetospheres of pulsars in binaries \citep{vigel06} and pulsar B's magnetosphere in particular \citep{arons04}. For instance, numerical simulations of A's relativistic wind interacting with B's dipole showed the bow-shock type structure formed around B \citep{arons04}. The shape of B's magnetosphere (see Fig. 2, \citet{arons04}) was very similar to that of the Earth's (see Fig. 12, \citet{tsyg02b}). However, these types of simulations are very computing power intensive, making the lightcurve fitting over the vast multi-dimensional parameter space unfeasible.

\citet{lyut05bp} showed that by considering non-trivial analytic models for B's magnetosphere one can reproduce the bright phases quite successfully. The approach was based on the assumption that pulsar B is intrinsically bright. However, the distortions due to the wind of A force the emission direction to be deflected away from our line of sight and render pulsar B unobservable. \citet{lyut05bp} used the similarities between the double pulsar and the Earth-Sun systems and developed a "stretched dipole" model for B's magnetosphere. The model accounted for the magnetosphere distortions due to the Chapman-Ferraro currents screening the dipole field from penetrating A's wind. Even though, the model neglected the effects from the other types of currents, it proved to be viable in terms of reproducing two distinct bright phases and secular changes in their position on the orbit. This feat is even more impressive considering that the simple circular emission beam centered on the polar field line was used and that the model was applied to B's "nightside" (side facing away from the pulsar A) magnetosphere only. Nevertheless, the success of "stretching" method in \citet{lyut05bp}, which is based on the simplified models of the Earth's magnetosphere \citep{stern87}, encourages the use of novel techniques for the development of more precise models of pulsar B's magnetosphere.

\subsection{Modified Tsyganenko's Model}

In the advanced three-dimensional model for pulsar B's magnetosphere, we used the planetary magnetosphere model by \citet{tsyg02a,tsyg02b} (T02). The T02 model is a data-based best-fit representation for the Earth's screened magnetosphere based on a large number of satellite observations. In addition to the Earth's dipole field, the model includes external magnetospheric sources such as the ring current, magnetotail current system, Chapman-Ferraro magnetopause currents, and the Region 1 and 2 Birkeland currents. The total magnetic field is comprised of modules as shown in equation \ref{earthfield}.

\begin{equation}
\label{earthfield}
B_{Tot}= B_{Dip} + B_{CF} + B_{RC} + B_{CT} + B_{BC} + B_{Int}
\end{equation}
where $B_{Dip}$ is the Earth's dipole field, while $B_{CF}$ is the field of the Chapman-Ferraro currents which flow in the magnetopause and confine this dipole inside the boundary. $B_{RC}$, $B_{CT}$, and $B_{BC}$ are the terms representing the contributions from the ring current, cross-tail current sheet, and the Birkeland currents, respectively. In order to ensure the full confinement of the total field within the magnetopause, each of these modules include the respective shielding field. The model also includes the interconnection field $B_{Int}$, which defines the amount of the solar wind's magnetic field that is penetrated through the magnetopause via reconnection. Thus, the fully screened magnetosphere corresponds to the fully "opaque" magnetopause with a zero interconnection field, while the highly resistive Dungey-type magnetosphere (Section \ref{sec:subpulsedrift}) corresponds to the fully "transparent" magnetopause.

The ring current is a principal source of the field deviation from the pure dipole at low altitudes. The equatorial drift of the pair plasma trapped in the magnetosphere generates the circular current that is coaxial with the Earth's dipole.
The exact nature of the Birkeland currents remains uncertain. However, they are assumed to flow into and out of the ionosphere along closed contours encircling the polar cap. At low altitudes, the fields are aligned with the diverging dipolar field lines, but then gradually stretch out at larger radial distances.
A third system is the cross-tail current flowing across the plasma sheet from dawn to dusk. This magnetotail current system is responsible for the stretched-out configuration of the tailward magnetosphere.

We used the GEOPACK library, numerical code for magnetospheric modeling developed by \citet{tsygurl}, with the appropriate modifications to match the properties of the double pulsar system (MTS model). The contribution of each term from the equation \ref{earthfield} is controlled by the parameters of the model, most of which are derived from observations. Instead of analyzing every current component in the T02 model separately, we manipulated the global input parameters of the code, which define the geometric structure of the magnetosphere. The shape and scale of the magnetosphere is controlled by the solar wind ram pressure and the dipole tilt only. Variations in the value of the ram pressure change the magnetosphere self-similarly. In the numerical model, the ram pressure is represented by the parameter PARMOD(1) and has units in nPa. PARMOD(2) represents the disturbance storm time (Dst) index. The Dst index is a measure of the size and strength of the ring current, which contributes to the overall field configuration in the inner magnetosphere.

The T02 model is designed in such a way that the structure of the magnetosphere within a standoff distance from the Earth has a very small dependence on the components of the interplanetary magnetic field (IMF). Additionally, T02 model does not allow the external wind field to spatially vary (e.g., magnetic field in a striped wind).

\begin{figure}
\centering
\includegraphics*[width=1.0\linewidth]{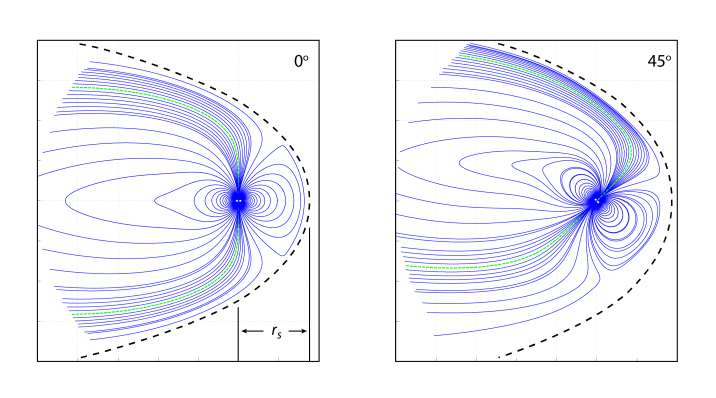}
\caption{T02 model-traced magnetic field lines of B for two different inclinations of the magnetic axis with respect to the line connecting the two pulsars. Field lines are plotted in solid blue, except the polar field lines which are plotted as green dashed lines. Black dashed line represents the theoretical model of the boundary. From \citet{per12}, the average standoff distance $r_{s}$ was assumed to be $\sim 4 \times 10^{9}$ cm.}
\label{fig:magnetosph_TS}
\end{figure}

We decided to reduce the wind-magnetosphere interaction essentially to a hydrodynamic confinement of B's magnetosphere by A's wind. Neglecting the IMF in the T02 model does not significantly change the goodness of fit of the bright phases, given that the effects of the striped wind from A should be smeared out on timescales much larger than the period of A. Therefore, due to the necessity of neglecting the IMF and the negligible impact on our results, we set the transverse components of the external field (PARMOD(3)=$B_{y}$ and PARMOD(4)=$B_{z}$) to zero.
To derive the values of the rest of T02 model parameters relevant to the double pulsar, we matched the boundary produced by the T02 code (see Figure \ref{fig:magnetosph_TS}) to the boundary produced by the theoretical model by \citep{gourg11, per12}.

The best fit values obtained from the visual fitting are PARMOD(1)=8 nPa for the solar wind ram pressure, PARMOD(2)=30 nT for the Dst index, and, by default, the zero transverse components of the IMF (PARMOD(3)=0 nT, PARMOD(4)=0 nT). Additionally, we had to rescale the stellar radius parameter $R0$ from 1 to 0.0026 since the standoff distance (the main characteristic length-scale of B's magnetosphere) produced by the model was about 10.4 stellar radii instead of 4000 stellar radii, which is the value assumed throughout this paper.

The values of the T02 parameters (PARMOD(1-4) and $R0$) obtained from the visual fitting of the magnetosphere boundaries are not supposed to be physically realistic; rather, they produce a magnetosphere with a shape and size (defined by the standoff distance) that matches the properties of the double pulsar. Moreover, there could be other successful fits since they are derived from the visual resemblance of the boundaries (see Figure \ref{fig:magnetosph_TS}). Nevertheless, using this particular set of parameter values suits our purpose of modeling an approximate size and shape of pulsar B's distorted magnetosphere without using large computational resources.
The structure of pulsar B's realistic magnetosphere is expected to be even more complicated. However, the T02 model proved to be robust enough to be adapted to the double pulsar.

In terms of magnetosphere geometry, the most notable differences between pulsar B and the Earth are in the spatial and temporal properties. For instance, the standoff distance for pulsar B's magnetosphere is about $4000$ neutron star radii \citep{lyut05dp,per12}. Whereas, in case of the Earth, the dayside magnetosphere extends only at $\sim 10-15$ stellar radii. We resolved this discrepancy by scaling the stellar radius parameter down to $0.0026$.

As for the differences in timescales, the rotation period of the Earth around its axis is about $31200$ times pulsar B's spin period. This translates into a light-cylinder radius for the Earth that is $31200$ times larger than pulsar B's, while the Earth's magnetosphere (standoff distance) is only $1.5$ times the size of B's magnetosphere.

\begin{figure*}
\centering
\includegraphics*[width=1.0\textwidth]{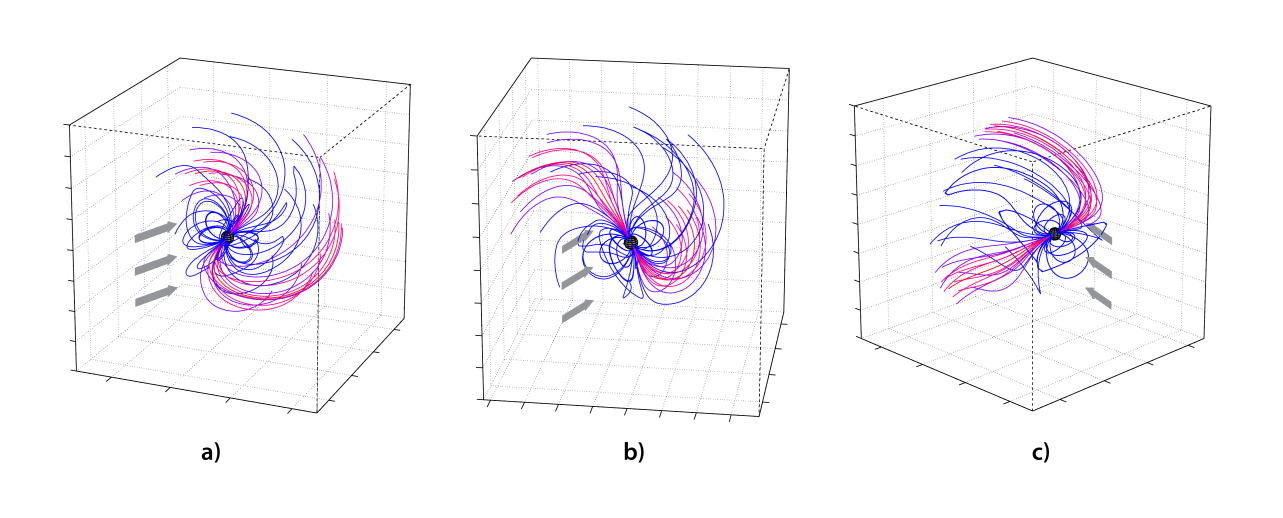}
\caption{Our MTS model-simulated structure of B's magnetosphere at three different spin phases. The field line retardation effect defined by the eq. \ref{sweepback} was imposed in order to simulate the effect of pulsar's rotation. The direction of A's wind is indicated by three grey arrows. Field lines originating from the region of the polar cap surrounding a pole is plotted in magenta.}
\label{fig:sweepback}
\end{figure*}

Since the relativistic distortions of the field lines and photon propagation direction are apparent only at a reasonable fraction of the light-cylinder radius \citep{shitov83,bltz91,romani95,Dyks04b,bai10}, these effects are absent in the Earth's magnetosphere. However, they become significant for the higher altitudes of pulsar B's magnetosphere. In order to account for these additional deflections, we added the retardation effect to the field lines generated by the T02 code. While the direction of the retardation is the opposite of the spin of pulsar B, its effect varies with the distance from the pulsar (see Fig. \ref{fig:sweepback}).

Various theoretical derivations yield different functional dependencies of the total deflection angle $\delta_{ret}$ on the normalized altitude $(\bar{r}=r/R_{LC})$. For instance, for an oblique rotator, \citet{shitov83} obtained $\delta_{ret} \sim \bar{r}^{3}$ by considering only the sweepback effect. \citet{bltz91} found that \citet{shitov83}'s estimate was underestimating the deflection magnitude by neglecting the aberration effects. Therefore, the resulted phase shift was of the first order in $\bar{r}$. \citet{Dyks04b} showed that relativistic distortions of the field depend not only on the altitude but also on the exact coordinates. They found that only the first few terms matter in a power series for $\delta_{ret}$ in $\bar{r}$. \citep{bai10} pointed out that most of the previous studies mistreated the transformations between the different reference frames and therefore, in general, resulted in incorrect expressions. Instead of leaning on a particular model, we approximated the retardation formula as a simple integer exponent of $\bar{r}$ with a scaling factor $\psi$ (see eq. \ref{sweepback}).

\begin{equation}
\label{sweepback}
\delta_{ret} = - \psi \left( \frac{r}{R_{LC}} \right)^{n}
\end{equation}

We tested the three most feasible values for $n$. Out of the tested values of $n = 1, 2,$ and $3$, a quadratic dependence consistently resulted in better fits.

\section{Simulation Setup} \label{sec:Method}

\begin{figure*}
\centering
\includegraphics*[width=0.8\linewidth]{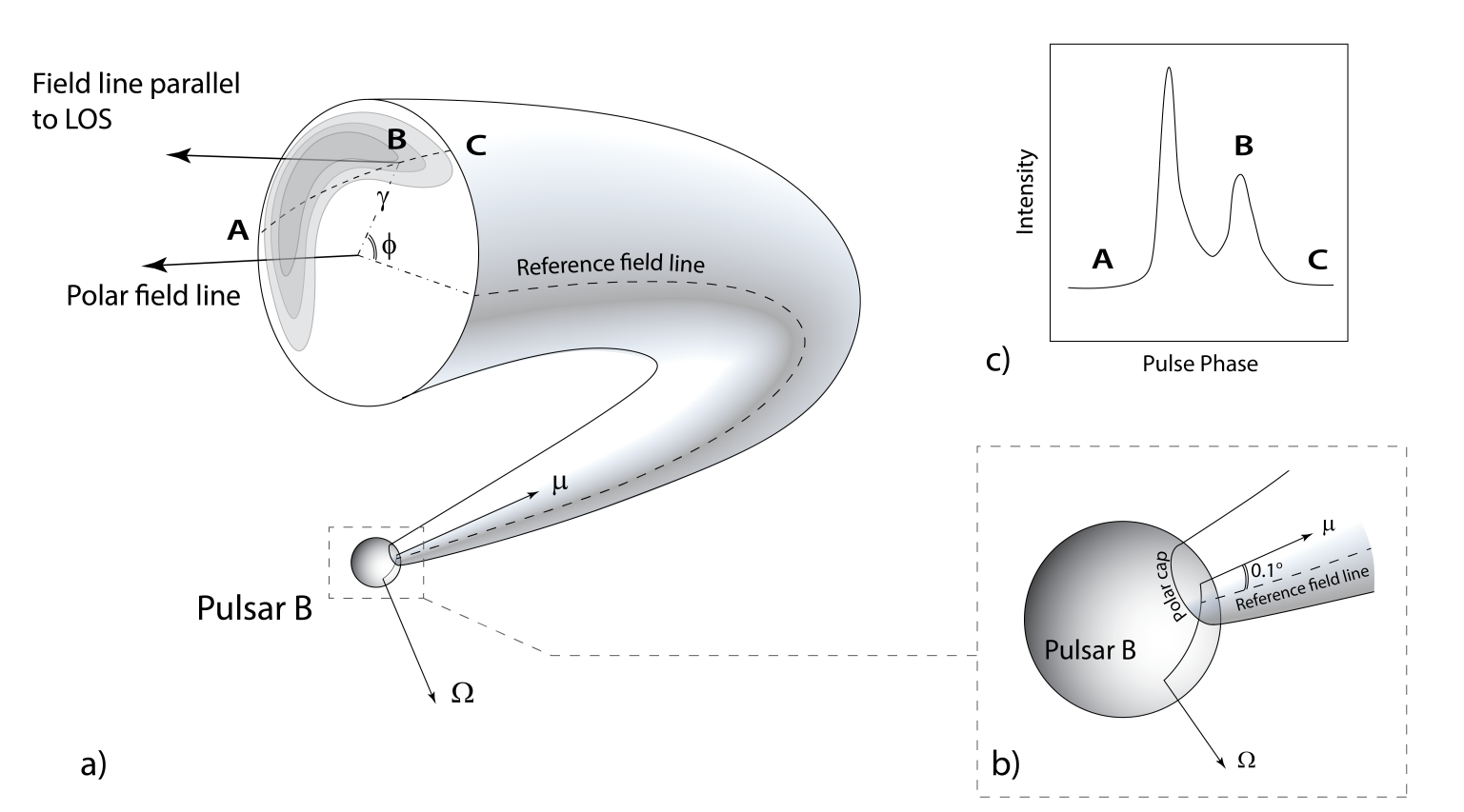}
\caption{Schematic view of one of the two distorted emission beams. It is assumed that the emission region is asymmetric and is centered on the polar field line (a). The reference field line (which is used to track the orientation of the emission region) originates close to the pole, and its footpoint lays on the shortest arc connecting the footpoints of the magnetic and spin axes (b). Two angles, $\gamma$ and $\phi$, fully describe the respective orientation of the LOS and the emission beam. $\gamma$ is an angle between the local direction of the polar field line (LDPFL) and the LOS. $\phi$ is an angle between the two planes, the plane containing the LOS and the LDPFL, and the plane containing the LDPFL and a local direction of the reference field line. Various cuts of the LOS through the emission beam yield various pulse profile morphologies. As an example, asymmetric two-hump profile (c) is generated along the ABC path of the LOS (a).}
\label{fig:los}
\end{figure*}

Our main goal is to determine the location and shape of the emission region by reproducing the modulation properties of B's radio lightcurves. The observed signatures of pulsar emission depend on the orientation of the emission direction (which, we assume, coincides with the local direction of the field lines in the emission region) with respect to our line of sight (LOS). However, for a precise modeling of the lightcurve, two components are necessary: the path of the LOS while passing through the emission beam and the intensity profile of the emission beam itself.
The passing path can be fully described by the two angles, $\gamma(t)$ and $\phi(t)$ (see Fig. \ref{fig:los} a)). $\gamma$ is a colatitude of the LOS with respect to the local direction of the polar field line (hereafter LDPFL), while $\phi$ is a rotation angle of the plane of the LOS and LDPFL with respect to the reference plane (see Fig. \ref{fig:los} a)). To define the reference plane, we follow the flux tube around the polar field line down to the NS surface. Therefore, we can use any plane that contains the LDPFL and a local direction of a field line whose footpoint is situated near the pole and corotates with the pulsar as a reference. In our case, we chose a reference plane that will coincide with the plane of magnetic and spin axes if traced down to the NS surface. Due to the asynchronous distortions of the different field lines with time, the direction of the reference plane changes non-trivially. We define the reference plane numerically by tracing the field line with a footpoint situated on the arc connecting the magnetic axis with the spin axis, and very close to the pole (see Fig. \ref{fig:los} b)).

In addition to the precise path of the LOS with respect to the LDPFL, in terms of the pair of angles $\gamma(t)$ and $\phi(t)$ as functions of time, knowing the intensity skymap I$(\gamma , \phi)$ (essentially an emission beam profile extended to the full sky) is required to model pulsar B's lightcurve. In the case of the filled circular emission beam, knowing only $\gamma$ is sufficient. However, \citet{per10} showed that this is not the case for the double pulsar. In section \ref{sec:EmRegion}, we discuss the detailed model of the beam structure, which describes how the emission is mapped on the plane of ($\gamma$,$\phi$). It should be noted that while $\phi$ ranges within $[0,2\pi]$, $\gamma$ only ranges from $0$ to $\pi$.

We calculated $\gamma(t)$ and $\phi(t)$ from modeling the time-resolved 3D structure of B's distorted magnetosphere and, in particular, field lines originating around the polar cap regions. We assume that the emission region is located close to the polar field lines (either or both); therefore, tracing only the field lines that surround the poles is sufficient.
From the intensity skymap (I$(\gamma , \phi)$) and the path of the LOS $(\gamma(t),\phi(t))$ we calculated the simulated lightcurve of B (see section \ref{sec:EmRegion}).

\begin{figure}
\centering
\includegraphics*[width=1.0\linewidth]{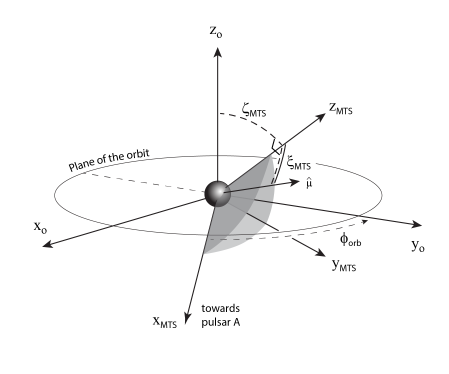}
\caption{Orientation of B's magnetic axis with respect to the wind direction. The origin of the coordinate system $(x_{\mathrm{MTS}},y_{\mathrm{MTS}},z_{\mathrm{MTS}})$ coincides with pulsar B, and the $x_{\mathrm{MTS}}$ axis points to the pulsar A (i.e., its antiparallel to the wind direction). In addition, $z_{\mathrm{MTS}}$ lies in the plane $x_{\mathrm{MTS}}-\mu$, making an acute angle with the magnetic axis $\mu$. The $(x_{\mathrm{MTS}},y_{\mathrm{MTS}},z_{\mathrm{MTS}})$ system is special since the GEOPACK code traces the field lines specifically in this frame. Produced distorted magnetosphere of B is symmetric with respect to the $x_{\mathrm{MTS}}-z_{\mathrm{MTS}}$ plane. Angle $\xi_{\mathrm{MTS}}$ defines the structure of the distortions; while $\zeta_{\mathrm{MTS}}$ and orbital phase $\phi_{orb}$ determine the rotations that are required for transforming the field line coordinates into the $(x_{o},y_{o},z_{o})$ system.}
\label{fig:axesMTS}
\end{figure}

An efficient way of parameterizing time is needed in order to simultaneously model the short-term (regular pulsed emission) and long-term modulations (changing orbital bright phases and pulse profile evolution) of B's radio emission in our computational grid.
One solution is to substitute time with rotational phases. This can be accomplished given that the state of B's magnetosphere can be fully described by three rotational phases: spin, orbital, and precession phases.
Spin phase is measured in the ($x_{s},y_{s},z_{s}$) coordinate system where B's spin axis coincides with the $z_{s}$ axis (see Fig. \ref{fig:pulsargeometry}). Orbital phase is measured clockwise from the ascending node (see Fig. \ref{fig:axesMTS}), while the precession phase $\phi_{prec}$ is calculated as an angle between the $x_{o}$ axis of the orbit fixed coordinate system and the plane of ($x_{s},z_{s}$). During one spin period, the orbital phase changes only by $\sim 0.24^{\circ}$. Considering that the systematic error is at least $1^{\circ}$ (due to the angular resolution of the generated field line database), we changed only the spin phase while keeping the orbital and precession phases constant for timescales of the order of B's spin period. Furthermore, B's spin precession phase changes only by $\sim 0.001^{\circ}$ over one orbital period. Therefore, on timescales of the order of B's orbital period, we changed only the spin and orbital phases and kept the precession phase constant. Finally, since all the processes considered here have timescales less than B's precession period of about 75 years, all three rotational phases were sufficient to treat the system's state as well as its evolution.
To simulate the lightcurves, instead of time-stepping, we performed a $1^{\circ}$ stepping of each phase parameter from $0^{\circ}$ to $359^{\circ}$ and calculated the intensity emitted towards the LOS. The resulting $1^{\circ}$ angular resolution in generated lightcurves is equivalent to 7 ms $(\approx P_{B}/360)$ in terms of the time resolution, which is much smaller than the average pulse width of about 80 ms.

The whole simulation process consists of three steps. As the first step, we traced B's magnetic field lines in the ($x_{MTS},y_{MTS},z_{MTS}$) system where the magnetic axis is tilted but stationary. In this system, the $x_{MTS}$ axis points towards pulsar A (antiparallel to the wind direction), while the $z_{MTS}$ axis lies in the plain that also contains $x_{MTS}$ and the magnetic axis of B. For every integer value of the tilt angle from $0^{\circ}$ to $180^{\circ}$, generated field line data was recorded in a database for later use (to save CPU time by avoiding repeated tracing of the same field lines).
In the second step, we calculated the necessary rotations to simulate the orbital motion and retardation effects and generate corresponding lighcurves using our emission beam model and the field line data stored in the database.
Finally, in the last step of our simulation we analyzed the lightcurves and fitted them to the observed data to estimate the properties of the emission region.

\section{Simulating The Lightcurves} \label{sec:Lightcurves}

The complexity of simulating the lightcurves of pulsar B is emphasized by the influence from its companion's wind. Since the magnitude and direction of the distortions depend on the respective orientation of the wind and the magnetosphere, lightcurve modeling requires knowing the field line structure for every combination of spin, orbital, and precession phases.

The distortions of B's magnetosphere are defined by the inclination of B's magnetic axis with respect to the wind direction. The inclination angle ($\pi/2 -\xi_{MTS}$) changes with time and therefore can be expressed with ($\phi_{spin},\phi_{orb},\phi_{prec}$), the spin, orbital, and precession phases, respectively.
In our simulations, pulsar B's magnetic field lines were traced in the wind-magnetosphere bound coordinate system ($x_{MTS},y_{MTS},z_{MTS}$) and transformed into the orbit-fixed ($x_{o},y_{o},z_{o}$) system (see Fig. \ref{fig:axesMTS}). This transformation was done by the superposition of the two rotations: R$(z_{o},\phi_{orb}-\pi/2)$ rotation by $\phi_{orb}-\pi/2$ around the $z_{o}$ axis, and R$(x_{o},-\zeta_{MTS})$ rotation by $-\zeta_{MTS}$ around the $x_{o}$ axis.

To calculate $\xi_{MTS}$ and $\zeta_{MTS}$, knowing the directions of the magnetic axis $\vec{\mu}$ and the wind (i.e. $x_{MTS}$) in the orbit fixed system was sufficient. The direction of $x_{MTS}$ is obtained by rotating the $x_{o}$ axis by $\pi/2 - \phi_{orb}$ around the $z_{o}$ axis. To find $\vec{\mu}$ in the ($x_{o},y_{o},z_{o}$) system, we transformed the magnetic axis $\vec{\mu}$ from the dipole's coordinate system (where $\vec{\mu}$ is along the $z_{dip}$) to the orbit-fixed system by applying the rotational operator R$(z, \phi_{prec})\otimes$R$(y, \theta)\otimes$R$(z, \phi_{spin})\otimes$R$(y, \alpha)$.

Additionally, we applied the retardation effect to the field lines, as defined by the eq. \ref{sweepback}. This was done by differentially rotating each point of the field line by angle $\delta_{ret}(r)$ around the spin axis $\Omega$ (i.e., the angle of rotation increases with distance from the pulsar).

Once obtained, the field line coordinates in the ($x_{o},y_{o},z_{o}$) system are used to calculate the ($\gamma, \phi$) paths of the LOS. By combining these ($\gamma, \phi$) paths with a model of the radio emission region, we simulated the lightcurves of pulsar B.

\subsection{Shape of the Emission Region} \label{sec:EmRegion}

The circular radio emission beam with a homogeneous intensity profile has been commonly used in pulsar studies.
However, a recent study of the evolution of the pulse profile widths in pulsar B's radio emission draws a more complex picture \citep{per10}, suggesting an elliptical horse-shoe shaped beam. Moreover, this type of beam structure is supported by the theoretical force-free models for an oblique rotator and, in particular, by a current distribution pattern in its polar cap region \citep{bai10,wang11}.

Due to the geodetic precession, the LOS passes through the emission beam each time at different distances from the polar field line. As a result, pulse profile widths and the separation of the peaks change (either by increasing or decreasing). By simulating the rate of this change, assuming the horse-shoe shaped emission beam, \citet{per10} estimated the ellipticity and the angular size of the beam.

\begin{figure}
\centering
\includegraphics[width=1.0\linewidth]{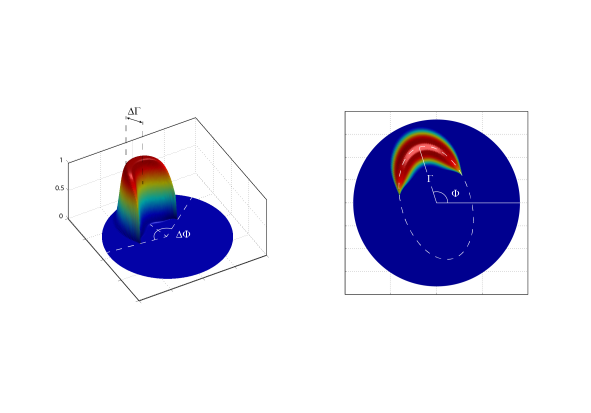}
\caption{Intensity profile of the emission beam, projected on the ($\gamma,\phi$) skymap. $\Gamma$ is a width of the emission beam; $\Phi$ is a rotation angle of the symmetry plane of the horse-shoe with respect to the orientation of the reference field line; $\Delta \Phi$ is a characteristic half-length of the arc (horse-shoe); $\Delta\Gamma$ is a maximum characteristic thickness of the horse-shoe.}
\label{fig:horseshoe}
\end{figure}

In our simulations, we also adopted a non-trivial structure of the emission beam. We assumed that radiation is emitted around the elliptical arc and that intensity has a super-Gaussian profile radially, as well as azimuthally. Visually, this is similar to an arc of a circular ring shrunk along the normal to the plane of symmetry (see Fig. \ref{fig:horseshoe}). The orientation of this symmetry plane depends on the spin, orbital, and precession phases as discussed in Section \ref{sec:Method}. The expression describing the normalized intensity $\mathrm{I}$ for the emission direction defined by the colatitude, $\gamma$, and longitude, $\phi$, is as follows:

 \begin{eqnarray}
 \label{beamprofile}
 \mathrm{I}=\exp\left(-16 \frac{\left( \tan \left(\frac{\phi-\Phi}{2}\right) \right)^{4}}{\Delta \Phi^{4}}\right) \exp \left(-\frac{\left(\gamma-\bar{\Gamma}\right)^{4}}{\Delta\bar{\Gamma}^{4}} \right) \\ \nonumber\\
 \bar{\Gamma}=\Gamma \frac{f}{\sqrt{\left( \sin \left(\phi-\Phi\right)\right)^{2}+\left( f \cos \left(\phi-\Phi\right)\right)^{2}}} \nonumber \\ \nonumber \\
 \Delta\bar{\Gamma}=\Delta\Gamma \exp \left(-\tan\left(\phi-\Phi\right)^{2}\right)\nonumber
 \end{eqnarray}

Here, $\Gamma$ is a width of the emission beam; $\bar{\Gamma}$ is an azimuth-dependent width; $f$ is a flatness factor (the ratio of the semi-minor and semi-major axes of the defining ellipse); $\Phi$ is a rotation angle of the symmetry plane of the horse-shoe with respect to the reference direction; $\Delta \Phi$ is a characteristic length of an arc (horse-shoe); $\Delta\Gamma$ is a maximum characteristic thickness of the horse-shoe; and $\Delta\bar{\Gamma}$ is an azimuth-dependent thickness. The visual representation of this model has a horse-shoe shape and is shown on Fig. \ref{fig:horseshoe}.

Both fully and partially covered cones can be reproduced by the model of the emission beam defined by eq. \ref{beamprofile}. The variable $\Delta \Phi$ defines to what extent the elliptical cone is covered. For instance, $\Delta \Phi = 180^{\circ}$ corresponds to the fully covered hollow cone emission beam centered on the polar field line. Additionally, we can model the hollow cone or filled emission beams by varying the parameters $\Gamma$ and $\Delta\Gamma$.
The normalized intensity $\mathrm{I}$ in eq. \ref{beamprofile} exhibits the required periodicity in terms of the azimuthal component $\phi$, and its profile has a super-Gaussian shape across the beam as well as around it. All described parameters of the emission beam ($\Phi,\Delta \Phi, \Gamma, \Delta \Gamma$, and $f$) were estimated later from the fitting of the simulated and observed lightcurves.

\subsection{Generated Lightcurves and Peak Intensity Maps} \label{sec:genLightcurves}

\begin{figure}
\centering
\includegraphics*[width=0.8\linewidth]{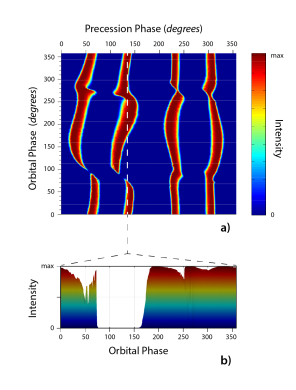}
\caption{Sample Peak Intensity Map (PIM) with a corresponding orbital lightcurve. Each vertical cut of the PIM corresponds to the orbital lightcurve at that precession phase (b). Therefore, PIM shows an evolution of the orbital lightcurve for the different precession phases (a). This particular lightcurve indicates the orbital modulation of the model emission.}
\label{fig:Lightcurve_2D}
\end{figure}

For each set of model parameters we generated a lightcurve which can be visualized in a number of ways.
In addition to the typical "intensity versus time" lightcurves, we can produce "intensity versus spin and orbital phases", "intensity versus spin and precession phases", and "intensity versus precession and orbital phases" maps by associating the intensity level of any point on the map with its brightness. Each of these maps is useful for reproducing and analyzing a particular property of B's radio emission. For instance, the "intensity versus spin and orbital phases" map is convenient in finding the orbital phase dependent changes in pulse profiles. The "intensity versus spin and precession phases" map is useful to simulate the secular evolution of pulse profiles (Fig. \ref{fig:PPE}), and the "intensity versus precession and orbital phases" map is most convenient for fitting the orbital modulation and its secular changes (Fig. \ref{fig:Lightcurve_2D} a)).

Since we are primarily interested in reproducing the bright phases and their evolution ("intensity versus precession and orbital phases"), we bin the simulated data in intervals with a duration of B's spin period in the following way: each bin is assigned an intensity level corresponding to the peak pulse intensity registered in that period. Due to the unequivocal relation between time and the set of ($\phi_{spin},\phi_{orb},\phi_{prec}$) rotational phases, we can also associate each bin with a certain orbital phase (while keeping the precession phase constant) in order to obtain the orbital lightcurve shown in Fig. \ref{fig:Lightcurve_2D} b). However, by varying the precession phase in the range of [$0^{\circ}, 359^{\circ}$] and aggregating the corresponding orbital lightcurves in a peak intensity map (hereafter PIM), we get a 2D pattern (see Fig. \ref{fig:Lightcurve_2D} a)) that allows us to identify the evolution of the bright phases.

Since it is possible to project a time interval on the precession phase interval and vice versa, each vertical cut on the PIM corresponds to the orbital lightcurve for that particular precession phase or date associated with that precession phase (see Fig. \ref{fig:Lightcurve_2D}). However, to define this projection, one needs to know the reference precession phase and the scaling factor, which in our case is the same as the geodetic precession rate of pulsar B: $\Omega_{prec} \approx 4.8^{\circ}$ yr$^{-1}$. At any moment ($t$), the precession phase can be expressed as $\phi_{prec}(t) = \phi^{0}_{prec} \pm \Omega_{prec} t$, with $\pm$ referring to the positive and negative directions of precession, respectively. Unlike the precession rate, we did not pre-assign a value to the reference precession phase ($\phi^{0}_{prec}$), but rather defined it from the fitting along with the precession direction (see section \ref{sec:Fitting}).

\section{Fitting and Analysis} \label{sec:Fitting}

We generated the peak intensity maps for single-pole and two-pole emission configurations for each set of the following 9 parameters: $\alpha$, the colatitude of the magnetic axis with respect to the spin axis; $\theta$, the inclination of the spin axis to the orbit normal ($z_{o}$);  $\psi$, the amplitude of the retardation angle; $R$, the emission height; $\Gamma$, the radius of the emission beam along the semi-major axis; $\Delta\Gamma$, the maximum characteristic thickness of the horseshoe; $\Phi$, the orientation of the beam's semi-major axis with respect to the reference plane (see Fig. \ref{fig:horseshoe}); $\Delta \Phi$, the characteristic angular half-width of the horseshoe arc; and $f$, the flatness factor (ratio of the semi-minor and semi-major axes of the defining ellipse of the horseshoe). For ($\alpha$,$\theta$,$\Phi$, and $\Delta \Phi$) we explored a full parameter sub-space, while we covered all values within a determined range of feasibility for the rest (see Table \ref{allparams}). The high number of free parameters involved in this problem does not permit the parameter sweep with small enough fixed step-sizes. Therefore, we used an iterative approach and continually refined the computational grid until we reached the desired level of the parameter estimate uncertainties (see Fig. \ref{fig:fits_all}).

\begin{figure}
\centering
\includegraphics*[width=1.0\linewidth]{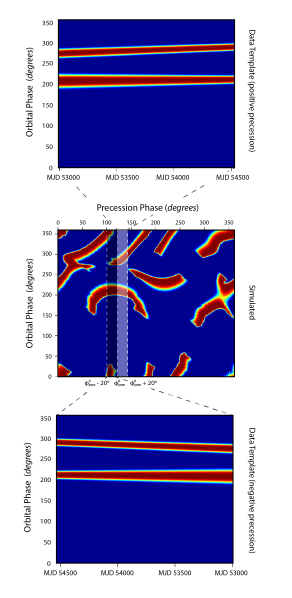}
\caption{Visualization of the fitting process. Two opposite orientations of the data template (upper and lower maps) are mapped on the precession phase intervals of $20^{\circ}$ span ($1500 \mathrm{days} \times 4.8^{\circ} \approx 20^{\circ} $). Then, the part of the simulated subimage from $\phi^{0}_{prec}$ to $(\phi^{0}_{prec} + 20^{\circ})$ (for the upper data template) is fitted with the data template by calculating their normalized cross-correlation coefficient. This is done for each integer value of $\phi^{0}_{prec}$ within [$0^{\circ}, 359^{\circ}$]. This way we can find the best-fit $\phi^{0}_{prec}$ with a maximum corresponding CC. We repeat these steps for every set of the model parameters values.}
\label{fig:fitting_method}
\end{figure}

\begin{figure*}
\centering
\includegraphics*[width=1.0\textwidth]{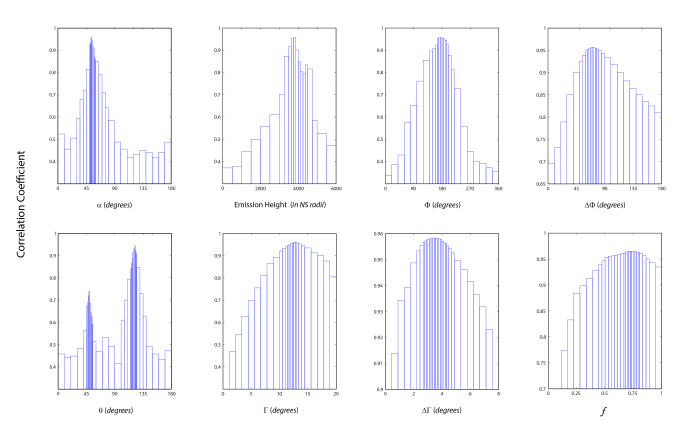}
\caption{Normalized cross-correlation coefficient (CC) for each of the eight model parameters: Emission Height, $\alpha$, $\theta$, $\Phi$, $\Delta\Phi$, $\Gamma $, $\Delta\Gamma$, and $f$. Higher CC indicates a better fit. All eight subplots show a peak in CC within the considered range, allowing us to determine the best-fit parameter value. Widths of the vertical bars correspond to the resolution of the simulation at a particular iteration. Therefore, the smallest width, in each subplot, represents the estimation uncertainty for that particular parameter. All best-fit parameter values and their uncertainties are given in Table \ref{allparams}.}
\label{fig:fits_all}
\end{figure*}

In order to compare the simulated evolution of the bright phases to the observational data by \citet{per10}, we structured the data in the same way as the generated PIMs (see the upper and lower plots on Fig. \ref{fig:fitting_method}). Thus, instead of its original "orbital phase versus MJD" form (see Fig. 10 in \citep{per10}), we projected the orbital modulation data onto the "orbital phase versus precession phase" map by the rule described in section \ref{sec:genLightcurves}. This map spans over the precession phase intervals of different lengths depending on the precession rate. The value of the geodetic precession rate that we used in our simulations ($4.8^{\circ}$ yr$^{-1}$) corresponds to the span of $20^{\circ}$. However, the initial (reference) precession phase, $\phi^{0}_{prec}$, is arbitrary and is defined from the fitting.
We also accounted for the negative precession (i.e., the precession axis antiparallel to the orbit normal $z_{o}$) by reversing the projection direction on the precession phase interval (see the lower plot in Fig. \ref{fig:fitting_method}).

We used the template matching via normalized cross-correlation \citep{gonz09} to compare the simulated and observed 2D peak intensity maps and find the best fit. This allows us to efficiently find a simulated PIM that contains the 2D pattern similar to the one in the data template (see Fig. \ref{fig:fitting_method}). In template matching algorithms, images are essentially treated as the $n \times m$ matrices with each element representing the brightness of the corresponding pixel. In our simulations, the PIMs are generated as $360 \times 360$ matrices. We characterized the goodness of fit by the correlation coefficient (see eq. \ref{CC}), which ranges from 0 to 1, with 1 representing perfect fit. The use of template matching allowed us to fit the location and width of the bright phases at any particular time (precession phase), but also for a continuous period of time (for the evolution of the bright phases).

\begin{equation}
\label{CC}
CC=\frac{\sum_{x,y}\left(F(x,y) - \bar{F} \right) \left(D(x,y) - \bar{D} \right)}{\sqrt{\sum_{x,y} \left(F(x,y) - \bar{F} \right)^{2}}\sqrt{\sum_{x,y} \left(D(x,y) - \bar{D} \right)^{2}}}
\end{equation}

Here, $CC$ is the cross-correlation coefficient of the template $D(x,y)$, with a subimage $F(x,y)$. $\bar{D}$ and $\bar{F}$ are the averages of the template and subimage, respectively.

We calculated the $CC$ for every set of the model parameters from the simulation grid. At every iteration, we selected the grid cell corresponding the maximum $CC$ and refined the grid around it. We repeated these steps until we reached the satisfactory level of parameter uncertainties. For each parameter, we plotted the maximum $CC$ for the values within a grid cell (see Fig. \ref{fig:fits_all}).

\section{Results} \label{sec:results}

We performed the fitting for both directions of the precession: along the orbital motion (prograde), and against it (retrograde). From the fitting, the prograde precession yielded a much worse fit than the retrograde; additionally, B's spin was in the same direction as its orbital motion.

The obtained best fit of the simulated PIM to the data template had the normalized correlation coefficient value of $CC \approx 0.96$ (out of 1). Some model parameters showed a larger variance of the $CC$ than others (see Fig. \ref{fig:fits_all}). For instance, the best-fit values of the emission height $R$, $\theta$, $\alpha$, and the radius $\Gamma$ and orientation $\Phi$ of the beam are more prominently defined than that of $\Delta \Gamma$, $\Delta \Phi$, and $f$. This is because the generated PIMs are less sensitive to the values of $\Delta \Gamma$, $\Delta \Phi$, and $f$. Nonetheless, all the model parameters show a peak $CC$ within the considered ranges of their values. Given that the $CC$ peaked within the range of values chosen, we presumed that the true best-fit lies within this range of values. We also defined the uncertainty interval as a smallest achieved size of the corresponding grid cell for each of the model parameters estimated through the fitting (see Table \ref{allparams}).

We estimated the emission height best-fit value as $3750 R_{NS}$, which corresponds to the peak $CC$ (see Fig. \ref{fig:fits_all}). This result was obtained using the model with a single emission height. However, it does not contradict the models with the emission generation within a narrow range of altitudes ($\pm 100 R_{NS}$).

Additionally, our best-fit value for the magnetic colatitude, $\alpha = 56^{\circ}$, is consistent with a previous estimate from the analysis of the secular changes in B's pulse profile widths by \citet{per10}. Moreover, all of the emission beam parameter values are in superb agreement with the ones obtained by \citet{per10} (see Table \ref{allparams}).

The estimated best-fit value of $\Phi = 180^{\circ}$ indicates that the footpoints of the beam field lines are situated around the pole directly across from the spin axis. In addition to being physically feasible, this result is consistent with the predictions of pulsar magnetosphere force-free models \citep{bai10,wang11}. The best-fit values for other parameters and their uncertainties are listed in Table \ref{allparams}.

\begin{table}
\caption{Best-fit estimates for the model parameters and their uncertainties}
\label{allparams}
\centering                          % used for centering table
\begin{tabular}{l c c r}        % centered columns (4 columns)
\\ \hline\hline  \\                % inserts double horizontal lines
Parameter                   & Best-Fit Value      & Range Tested      & Min. Grid Size \\\\
\hline  \\                      % inserts single horizontal line
Emission Height            & $3750 R_{NS}$       & $[0, 6000 R_{NS}]$              & $50 R_{NS}$\\
$\alpha$                   & $56^{\circ}$        & $[0^{\circ}, 180^{\circ}]$      & $1^{\circ}$\\
$\theta$                   & $122^{\circ}$       & $[0^{\circ}, 180^{\circ}]$      & $1^{\circ}$\\
$\Gamma$                   & $13.2^{\circ}$      & $[1^{\circ}, 20^{\circ}]$       & $0.2^{\circ}$\\
$\Delta\Gamma$             & $3.5^{\circ}$       & $[0.5^{\circ}, 8^{\circ}]$      & $0.05^{\circ}$\\
$\Phi$                     & $180^{\circ}$       & $[0^{\circ}, 359^{\circ}]$      & $5^{\circ}$\\
$\Delta\Phi$               & $72^{\circ}$        & $[0^{\circ}, 180^{\circ}]$      & $2.5^{\circ}$\\
$f$                        & $0.75$              & $[0.1, 1.0]$                    & $0.025$\\
$\phi^{0}_{prec}$          & $122^{\circ}$       & $[0^{\circ}, 359^{\circ}]$      & $1^{\circ}$\\
$\psi$                     & $2.2$               & $[-4, 4]$                     & $0.2$\\\\
\hline                                   %inserts single line
\end{tabular}
\end{table}

\begin{figure}
\centering
\includegraphics*[width=1.0\linewidth]{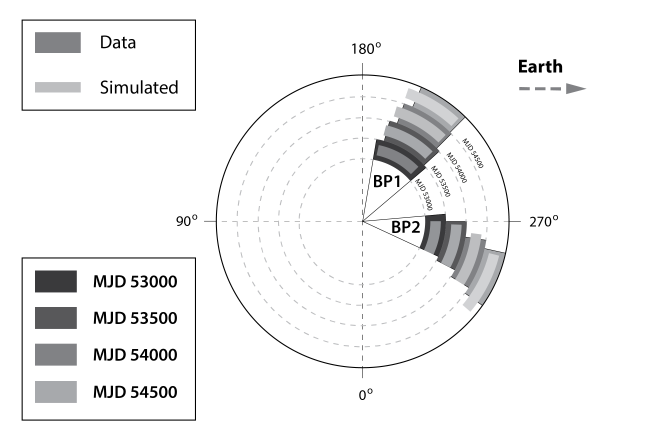}
\caption{Schematic view of the evolution of the orbital bright phases with overlaid best-fit results of our simulations. Different shades of grey represent different epochs of the observations. Durations and phases of the bright phases changed gradually with time as shown on the figure. }
\label{fig:BPResults}
\end{figure}

The MTS model reproduced the location and extent of the bright phases with up to one degree precision when fitted to the data from any particular epoch. However, the best fit obtained from the simultaneous fitting over the range of dates showed larger deviations from the observational data. The comparison between the simulated best-fit and observed bright phases is shown in Fig.\ref{fig:BPResults}.

\subsection{Reproducing B's Disappearance with a Null-charge Surface Cutoff} \label{sec:cutoff}

\begin{figure}
\centering
\includegraphics*[width=1.0\linewidth]{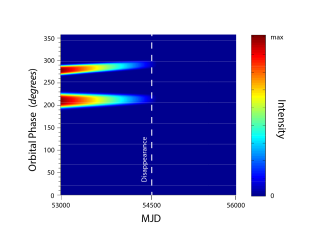}
\caption{Peak Intensity Map representation of the observational data for pulsar B's bright orbital phases. Our approximated disappearance time of MJD 54500 is plotted as a vertical dashed line.}
\label{fig:data_disappearance}
\end{figure}

\begin{figure}
\centering
\includegraphics*[width=1.0\linewidth]{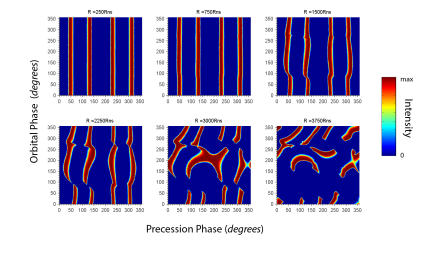}
\caption{Simulated Peak Intensity Maps for different altitudes of the emission region. No cutoff has been applied. The parameter values used in these simulations correspond to the best-fit estimates: $\alpha = 56^{\circ}$, $\theta = 122^{\circ}$; Beam parameters: $\Gamma = 13.2^{\circ}$, $\Delta\Gamma = 3.5^{\circ}$, $\Phi = 180^{\circ}$, $\Delta\Phi = 72^{\circ}$, $f = 0.75$.}
\label{fig:lightcurves}
\end{figure}

In addition to the change in the longitude and widths of pulsar B's bright phases, its radio emission gradually decayed over the period of [MJD 53000 - MJD 54500] and eventually disappeared for at least 1500 days (private comm. B. Perera). In terms of PIMs, this is equivalent to zero observed intensity over the precession phase interval with a span of $\sim 20^{\circ}$ (see Fig. \ref{fig:data_disappearance}). This trend cannot be reproduced without additional tweaking of the model (Fig. \ref{fig:lightcurves}).

In conventional models of precessing binary pulsars, spin axis precession causes the emission direction to miss the observer and essentially disappear \citep{weis89,kram98}. This is the case for a nearly dipolar magnetosphere with minor distortions (see the upper row in Fig.\ref{fig:lightcurves}). However, despite reproducing the disappearance of B, lower emission altitudes cannot account for the significant orbital modulation of the radio emission. On the other hand, at higher altitudes (see the lower row in Fig.\ref{fig:lightcurves}), distortions are strong enough to cause significant orbital modulation. Due to the estimated high emission altitude in pulsar B (and thus significant distortions of the field lines), the spin axis precession is not sufficient to keep the emission beam averted from the LOS over the course of pulsar's rotation around its spin axis. Thus, B is always observable, which contradicts past documentation of B's undetectability. Regardless of the precession phase, the wind from A pushes the emission beam back in the observer's direction (at certain spin phases).

We discovered that applying an additional cutoff, defined by the eq. \ref{cutoff}, to an intensity of B's radio emission (eq. \ref{fullinten}) provides satisfactory results on simultaneously reproducing the disappearance of pulsar B and the evolution of its bright phases. The total intensity after applying the cutoff is:

\begin{equation}
\label{fullinten}
\overline{\mathrm{I}}=\mathrm{I}_{1} \mathrm{U}(\eta_{\Omega \mathrm{B},1}) + \mathrm{I}_{2} \mathrm{U}(\eta_{\Omega \mathrm{B},2})
\end{equation}

\begin{figure}
\centering
\includegraphics*[width=0.8\linewidth]{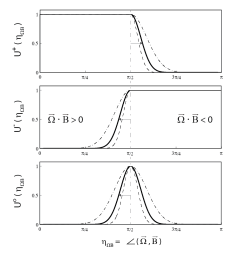}
\caption{Functional representation of the cutoffs imposed on the radio emission in order to reproduce B's disappearance. Using $\mathrm{U}^{+}$ is equivalent of allowing the radio emission only when $\Omega \cdot \mathrm{B} \geq 0$, where $\mathrm{B}$ is a local magnetic field and $\Omega$ is the spin axis. Conversely, $\mathrm{U}^{-}$ is equivalent of allowing the radio emission generation only when $\Omega \cdot \mathrm{B} \leq 0$. $\mathrm{U}^{o}$ is equivalent of allowing the radio emission generation only near the null-charge surface ($\Omega \cdot \mathrm{B} \sim 0$). Dash-dotted lines correspond to the characteristic half-width $\epsilon = 0.4$; dash lines correspond to $\epsilon = 0.17$; Solid line corresponds to the best-fit value of $\epsilon = 0.25$ $(\sim 14^{\circ})$.}
\label{fig:cutoffs}
\end{figure}

Here, $\mathrm{I}_{1}$ and $\mathrm{I}_{2}$ are the intensity contributions from each of the emission regions (assuming a two-pole configuration). $\eta_{\Omega \mathrm{B}}$ is the angle between the spin axis and the local magnetic field. We parameterized the cutoff with a function $\mathrm{U}(\eta_{\Omega \mathrm{B}})$, which acts as a filter, permitting non-zero responses only for certain values of the angle $\eta_{\Omega \mathrm{B}}$ (see Fig. \ref{fig:cutoffs}). We tested three forms for $\mathrm{U}(\eta)$:

\begin{eqnarray}
\label{cutoff}
\mathrm{U}^{0}(\eta) = \exp\left({-\frac{\left(\eta - \pi/2 \right)^{2}}{\epsilon^{2} }}\right)\\
\mathrm{U}^{+}(\eta) = \exp\left({-\frac{\left(\eta - \pi/2 \right)^{2}}{\epsilon^{2} } \mathrm{Heaviside}\left(\eta - \pi/2 \right)}\right) \\
\mathrm{U}^{-}(\eta) = \exp\left({-\frac{\left(\eta - \pi/2 \right)^{2}}{\epsilon^{2} } \mathrm{Heaviside}\left(\pi/2  - \eta \right)}\right)
\end{eqnarray}

For all three forms of the filter $U$ in eq. \ref{cutoff}, the transition in the response happens at $\eta_{\Omega \mathrm{B}} = \pi/2$ (see Fig. \ref{fig:cutoffs}), which corresponds to the null-charge surface.
$\mathrm{U}^{0}(\eta)$ is a Gaussian-like filter with a characteristic half-width $\epsilon$. $\mathrm{U}^{+}(\eta)$ and $\mathrm{U}^{-}(\eta)$ have a stepfunction-like behavior. $\mathrm{U}^{+}(\eta)$ matches $\mathrm{U}^{0}(\eta)$ for the values of $\eta \geq \pi/2$, while $\mathrm{U}^{-}(\eta)$ matches $\mathrm{U}^{0}(\eta)$ for the values of $\eta \leq \pi/2$.

Using the cutoff $\mathrm{U}^{0}(\eta_{\Omega \mathrm{B}})$ is equivalent to imposing the condition that the radio emission is generated only within a small angle from the null-charge surface ($\Omega \cdot \mathrm{B} \sim 0$). In addition to the angular spread permitted by $\mathrm{U}^{0}(\eta_{\Omega \mathrm{B}})$, $\mathrm{U}^{+}(\eta_{\Omega \mathrm{B}})$ extends the possible emission generation sites to the values of $\eta \leq \pi/2$ (i.e., $\Omega \cdot \mathrm{B} \geq 0$). Whereas $\mathrm{U}^{-}(\eta_{\Omega \mathrm{B}})$ extends them to $\eta \geq \pi/2$ (i.e., $\Omega \cdot \mathrm{B} \leq 0$).

\begin{figure}
\centering
\includegraphics*[width=1.0\linewidth]{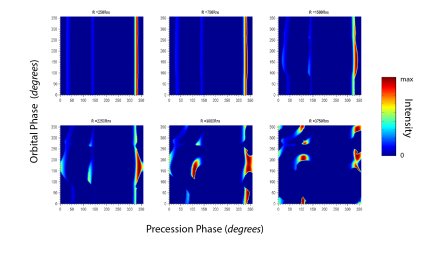}
\caption{Same as the Fig.\ref{fig:lightcurves} but with an imposed cutoff $\mathrm{U}^{o}$. Thus, only the emission generated within $14^{\circ}$ of the null-charge surface is registered.}
\label{fig:lightcurves_cutoff_ns}
\end{figure}

\begin{figure}
\centering
\includegraphics*[width=1.0\linewidth]{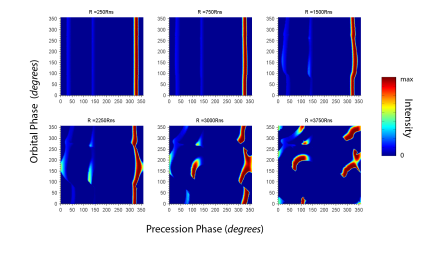}
\caption{Same as the Fig.\ref{fig:lightcurves} but with an imposed cutoff $\mathrm{U}^{-}$. Thus, mainly the emission generated within the $(\Omega \cdot \mathrm{B} \leq 0)$ region is registered.}
\label{fig:lightcurves_cutoff_minus}
\end{figure}

For all three types of the cutoffs, we repeated the fitting of the simulated peak intensity maps to the revised data template (with the added disappearance of B's bright phases, see Fig. \ref{fig:data_disappearance}). The implementation of $\mathrm{U}^{0}$ and $\mathrm{U}^{-}$ into the MTS model allowed us to reproduce the disappearance of pulsar B (see Fig. \ref{fig:lightcurves_cutoff_minus} and \ref{fig:lightcurves_cutoff_ns}). However, applying the $\mathrm{U}^{+}$ cutoff to the emission model did not yield a reasonable fit. The value of $14^{\circ}$ for the half-width of the cutoff $\epsilon$ produced the best fit with the observed rate of B's disappearance. Additionally, all other parameter estimates were consistent with the results of the model without a cutoff.

Given that the $\mathrm{U}^{0}$ cutoff produced a realistic disappearance rate, the observed emission came from the near null-charge surface layer of B's magnetosphere. Moreover, if the observed emission was generated within the $\Omega \cdot \mathrm{B} \leq 0$ region then $\mathrm{U}^{-}$ would not be able to reproduce the disappearance of B, since $\mathrm{U}^{-}(\eta_{\Omega \mathrm{B}}  \leq 0) = 1$. However, Fig. \ref{fig:lightcurves_cutoff_minus} shows that within the interval of the precession phases corresponding to the epoch of B's "visibility" ($\sim$ MJD 53000 - MJD 54500), the PIMs generated with the $\mathrm{U}^{0}$ and $\mathrm{U}^{-}$ cutoffs match each other exactly. Therefore, B's radio emission observed during [MJD 53000 - MJD 54500] came primarily from the near null-charge surface layer of the $\Omega \cdot \mathrm{B} \geq 0$ region.

Finding the exact nature of the processes responsible for these cutoffs is beyond the scope of this work. However, it should be noted that the adjacent layers of the null-charge surface host various magnetospheric gaps, which represent the spatially limited regions of particle acceleration \citep{usov99}.

Using $\mathrm{U}^{+}$ is equivalent of allowing the radio emission only when $\Omega \cdot \mathrm{B} \geq 0$, where $\mathrm{B}$ is a local magnetic field and $\Omega$ is the spin axis. Conversely, $\mathrm{U}^{-}$ is equivalent of allowing the radio emission generation only when $\Omega \cdot \mathrm{B} \leq 0$. $\mathrm{U}^{o}$ is equivalent of allowing the radio emission generation only near the null-charge surface ($\Omega \cdot \mathrm{B} \sim 0$).

\begin{figure}
\centering
\includegraphics*[width=1.0\linewidth]{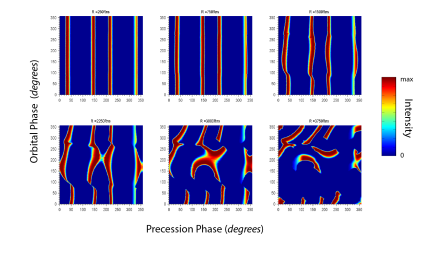}
\caption{Same as the Fig.\ref{fig:lightcurves} but with an imposed cutoff $\mathrm{U}^{+}$. Thus, mainly the emission generated within the $(\Omega \cdot \mathrm{B} \geq 0)$ region is registered.}
\label{fig:lightcurves_cutoff_plus}
\end{figure}

\section{Implications} \label{sec:Implications}

\subsection{Reproducing the Pulse Profile Evolution} \label{sec:PPE}

In addition to the orbital modulation of the radio emission, pulsar B exhibited the evolution of its pulse profile from single to double peak. Studying this evolution allowed \citet{per10} to determine the shape of the emission beam as well as to estimate the geometry of B. We adopted the general morphology of the emission beam obtained by \citet{per10} and estimated its detailed characteristics through the fitting of the widths of the bright phases. Successfully simulating the pulse profile evolution provided an additional test for our model.

\begin{figure}
\centering
\includegraphics*[width=1.0\linewidth]{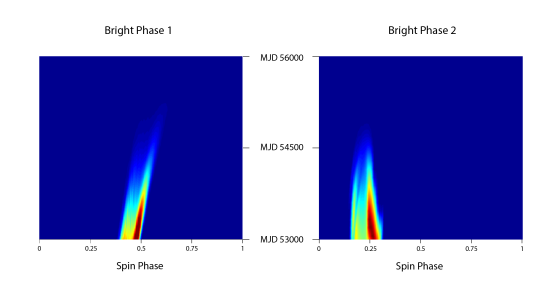}
\caption{Simulated secular evolution of B's pulse profiles for the BP1 (left) and BP2 (right).}
\label{fig:PPE}
\end{figure}

In order to reproduce the observed pulse profile evolution in B's radio emission, we assumed that the model parameters were equal to the estimated best-fit values (obtained in the previous sections) and integrated the simulated emission intensity over the orbital phases corresponding to BP1 and BP2. As a result, the signature of the evolution from a single peak to a double peak pulse profile, though weak, can still be seen in the Fig.\ref{fig:PPE} for both BP1 and BP2.

\subsection{Reappearance of PSR J0737-3039B} \label{sec:Reappearance}

\begin{figure}
\centering
\includegraphics*[width=1.0\linewidth]{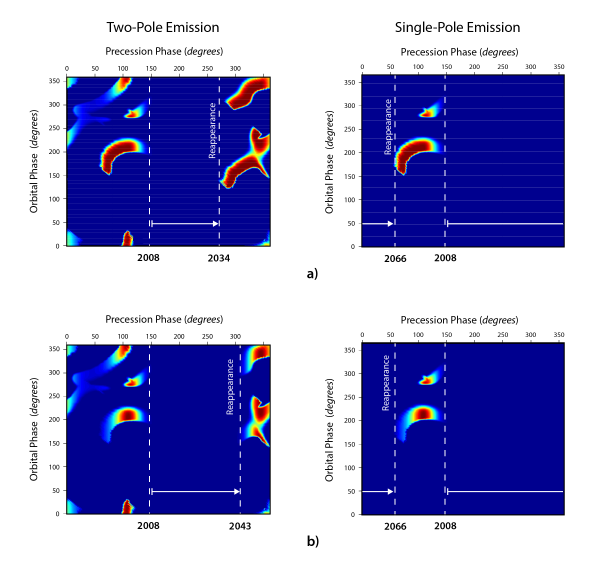}
\caption{Predicted reappearance of pulsar B for two-pole and single-pole emission geometries. a) Pulsar B's best fit Peak Intensity Maps (PIM) with a condition that the radio emission is only generated if $\Omega \cdot \mathrm{B} \leq 0$. Here, B is a local magnetic field. b) B's PIM with a condition that the radio emission is only generated if $\Omega \cdot \mathrm{B} \sim 0$. Future observations will determine which configuration is at play.}
\label{fig:Reappearance}
\end{figure}

According to our simulation results, the reappearance time of B depends on some aspects of the theory of pulsar radio emission, along with the size and shape of the emission region and the model of B's magnetosphere. Namely, the reappearance time of B differs based on whether the radio emission can be generated only in one particular hemisphere of the pulsar or both. Additionally, it is also influenced by whether or not the radio emission is generated near the null-charge surface only. For the most part, the model yielded different answers for each of the four arrangements (see Fig. \ref{fig:Reappearance}). In the case of two-pole emission, the reappearance is supposed to happen in the year $\sim 2034$ for the $(\Omega \cdot \mathrm{B} \leq 0)$ emission and in the year $\sim 2043$ for the near null-surface emission region. For both of these cases however, we expect a similar reappearance date of year $\sim 2066$ if the emission is from a single pole. Nonetheless, it will be possible to distinguish between the two locations of the emission generation by analyzing the growth of B's flux density shortly after its reappearance (see the two plots on the right in Fig. \ref{fig:Reappearance}). Moreover, in case of a single-pole emission, pulsar B is expected to exhibit only one bright phase when it reappears.

\subsection{Reducing Degeneracies in the Double Pulsar Geometry} \label{sec:Degeneracies}

\begin{figure}
\centering
\includegraphics*[width=1.0\linewidth]{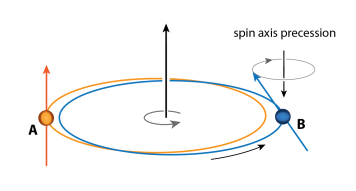}
\caption{Configuration of the spin-orbital angular momenta of PSR J0737-3039A/B. Respective orientation of B's spin axis and orbital angular momentum were drawn according to our best-fit results. A's spin axis direction was acquired from \citep{ferd2013}.}
\label{fig:Degeneracies}
\end{figure}

Our model of B's distorted magnetosphere allowed us to derive the relative orientations of all three axes, B's spin, the orbital angular momentum, and the precession axes, independently. However, the orientation of the angular momentum with respect to the plane of the sky remains uncertain. Due to the insufficient angular resolution of our computational grid, we were unable to determine the sign of the inclination angle $i$ (see Fig. \ref{fig:pulsargeometry}). Nonetheless, B's spin and spin axis precession directions were uniquely defined with respect to the orbital motion. In the coordinate system ($x_{o}, y_{o},z_{o}$) (see Fig. \ref{fig:pulsargeometry}), the angular momentum is antiparallel to the $z_{o}$ axis. Thus, the orbital motion is clockwise when looking from the top. The best-fit value of the retardation parameter $\psi$ and its sign (both positive and negative values of $\psi$ were tested to account for the direction of the spin) indicates that B's spin axis is in the same hemisphere as the orbital angular momentum, with respect to the orbital plane (see Fig. \ref{fig:Degeneracies}). Therefore, the colatitude of the spin axis with respect to the orbital angular momentum is effectively $58^{\circ}$ $(180^{\circ}-\theta)$.

We also estimated the direction and the reference phase $\phi^{0}_{prec} \approx 122^{\circ}$ (i.e., $\phi_{prec}$ corresponding to MJD 53000) of the spin axis precession. Simulation results implied that the direction of B's spin axis precession is the opposite of the direction of the orbital motion. One could argue that this is against theoretical predictions. According to general relativity, the spin axis precesses around the total angular momentum \citep{damur92}. In the double pulsar, however, the orbital angular momentum makes up more than 99.9\% of the total angular momentum \citep{kram06}. Therefore, instead of being antiparallel, the general relativity predicts the precession axis and the orbital angular momentum to be nearly parallel.

\section{Estimating the Emission Height Using the Drifting Subpulses} \label{sec:subpulsedrift}

Alternatively, we use the peculiar nature of the subpulse drift in B's radio emission to estimate the emission height and study the properties of pulsar wind. \citet{mcl04b} argues that the reason for the observed modulation could be the high frequency ($\sim 44\,\mathrm{Hz}$) changes in the emission direction due to the influence of A's radiation field. We suggest that such changes should be attributed to the field line distortions caused by the reconnection of the field lines of B's dipole with the magnetic field lines in the companion's striped wind. The existence of such pulsar winds is supported by theory; \citet{bog99} showed that pulsars should expel the striped winds with periodically varying radial profiles of density and pressure carrying a magnetic field with similar structure.

In the case of strongly magnetized wind and magnetosphere, the distortions depend on the relative
strengths of the magnetic fields and thus on the distance from the neutron star. Close to the pulsar, the deflections of the field lines are diminishingly small and gradually increase outwards. This means that there is a certain height above which the distortion is strong enough to deflect the field lines encompassing the emission region more than an angular size of the latter, pushing it out of the line-of-sight. Therefore, an observer will detect different radiation signatures of the distorted magnetosphere depending on the location of the radio emission region. By studying these signatures, we can deduce the magnitude of the distortions and thus the location of the emission region.
However, studying the effects of distortions on the observed radio emission requires a development of a time-resolved 3D model of wind-magnetosphere interaction.

\subsection{Dungey-type Model of B's Magnetosphere}

The nature of the reconnection process between the pulsar and wind magnetic fields is very similar to what Dungey proposed for planetary magnetospheres. \citet{dungey61} model states that the interplanetary magnetic field (IMF) may become reconnected with the terrestrial field along the day-side magnetopause, which results in a distortion of the higher altitude regions of the inner magnetosphere.

\begin{figure}
\centering
\includegraphics*[width=0.8\linewidth]{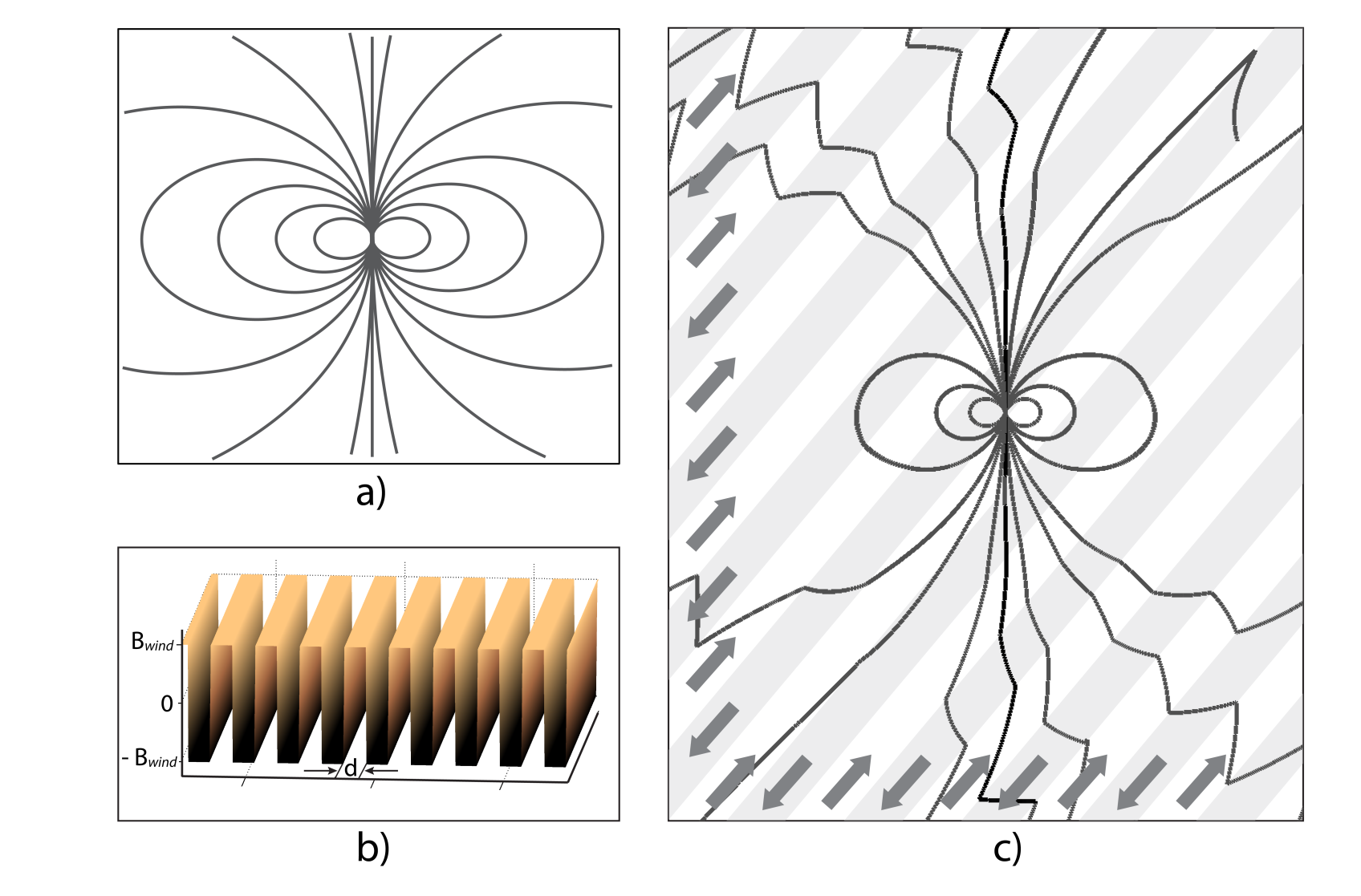}
\caption{The Dungey-type field with a striped wind. (a) A simple dipole field projected on 2D; (b) The distribution of the magnetic flux density in the striped wind. $d = c \times P_{A}/2 = 3.3 \times 10^{8} \mathrm{cm}$ is the width of each stripe; (c) Dungey-type field. Result of our numerical simulation based on a linear superposition of dipole and wind's striped fields. Shaded stripes represent the regions with the magnetic field of a uniform absolute value but opposite polarity. The small arrows at the borders mark the direction of the magnetic field for each stripe.}
\label{fig:dungey}
\end{figure}

However, the spatial and temporal properties of these processes are different for the Earth and the double pulsar. In the case of the Sun, field changes occur on scales much larger than the size of the magnetosphere, so at each
moment the magnetosphere is subject to a nearly constant external magnetic field. This results in the southward-northward asymmetry seen in the Earth's magnetosphere. In the case of the double pulsar, the direction of the field in the pulsar wind changes on a distance equal to the half period of pulsar A multiplied by the speed of light, $d = c \times P_{A}/2 = 3.3 \times 10^{8} \mathrm{cm}$. This creates a striped wind structure (Fig. \ref{fig:dungey} (b)) with a length-scale, an order of magnitude smaller than the size of B's magnetosphere \citep{lyut05dp}. This leads to the conclusion that instead of exhibiting the southward-northward asymmetry the structure of B's magnetosphere should be rather jittery (Fig. \ref{fig:dungey} (c) ).

In order to model the high frequency ($44\,\mathrm{Hz}$) distortions of B's magnetic field lines, we developed a simplified model using the prescription devised by \citet{dungey61} and \citet{forb71} for planetary magnetospheres.
The latter neglected the dynamics of the reconnection processes and modeled the
Earth's magnetosphere as a linear superposition of two magnetic fields: the Earth's closed dipole field
and the solar wind's uniform field. Following this approach while keeping in mind the spatial scales of our problem, we can represent B's magnetosphere as a simple addition of the pulsar's dipole field and the wind's striped field (Fig. \ref{fig:dungey} (a) and (b)).

From the dawn of pulsar physics, the simple dipole representation of the neutron star magnetosphere has been successfully used to understand not only simple phenomena but also a number of complex ones. Similarly, it is sufficient to assume a pure dipole as an intrinsic field of pulsar B, for the purpose of modeling the distortions.

We carry out the simulation in the frame of the static dipole with a magnetic axis along the $\hat{z}$ axis. The components of the dipole in this frame are:

\begin{eqnarray}
\label{dipfield}
B_{dip}^{x}&=&\frac{\mu_{0}M}{4\pi}\frac{x z}{(x^{2}+y^{2}+z^{2})^{5/2}}\nonumber \\
B_{dip}^{y}&=&\frac{\mu_{0}M}{4\pi}\frac{y z}{(x^{2}+y^{2}+z^{2})^{5/2}}\\
B_{dip}^{z}&=&\frac{\mu_{0}M}{4\pi}\left(\frac{3z^{2}}{(x^{2}+y^{2}+z^{2})^{5/2}}-\frac{1}{(x^{2}+y^{2}+z^{2})^{3/2}}\right)\nonumber
\end{eqnarray}

Here, $M=\frac{2\pi B_{NS} R_{NS}^{3}}{\mu_{0}}$ is a magnetic moment of the neutron star. We assume that the radius of the neutron star $R_{NS}$ is $10^{6}\mathrm{cm}$ and borrow the value of the surface magnetic field $B_{NS} \approx 6.4 \times 10^{11} \mathrm{G}$ from \citep{per12}.

\begin{figure*}
\centering
\includegraphics*[width=0.8\textwidth]{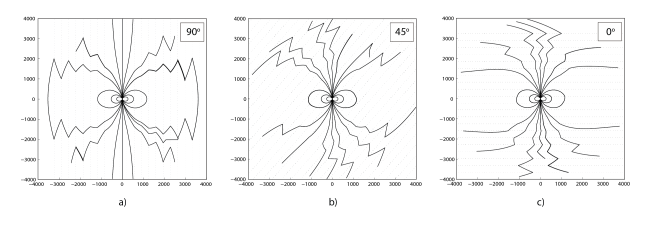}
\caption{The Dungey-type field with a striped wind. Overall structure of B's magnetosphere for different incident angles of A's wind (the line connecting the two pulsars). An angle between A's wind and B's magnetic axis equals to $90^{\circ}$, $45^{\circ}$, and $0^{\circ}$, for (a), (b) and (c) respectively. The strengths of the dipole's and winds magnetic fields were assumed to be equal at $4000 R_{NS}$ along the dipole equator.}
\label{fig:dungey3}
\end{figure*}

In the double pulsar, we assume a toroidal wind from A is hitting pulsar B with a direction of the magnetic flux density vector perpendicular to the line connecting the two pulsars \citep{bog99}. However, in the vicinity of pulsar B (length-scales much smaller than orbital radius), the curvature of the torus is negligible. Therefore, it is sufficient to consider the plane stripes propagating through the magnetosphere. Moreover, the model field is set to be coplanar to the orbital plane and to flip its direction between two consecutive stripes while staying uniform in absolute value, with an exception of the discontinuities between the stripes (Fig. \ref{fig:dungey3}).

We assume that the magnetic field in the stripped wind consists of the regions of constant magnetic field,
measuring half a wavelength, separated by a current sheet from the regions of opposite polarity. We can model this type of field by summing the step-functions of opposite signs and shifted phases. This will satisfy our requirements for the striped wind, by experiencing periodic jumps along one coordinate ($\bar{x}$) (which is also its propagation direction) while staying homogeneous along the other two ($\bar{y}$ and $\bar{z}$) (Fig. \ref{fig:dungey} b)). Components of the magnetic flux density at the time $t$ for such field in the frame of the pulsar wind can be expressed as:

\begin{eqnarray}
\label{windfield}
B_{wind}^{\bar{x}}(t)&=&B_{wind}^{0}\left(\sum_{n=0}\frac{2\,(-1)^{n}}{1+exp(-2k(\bar{x}-nd-vt))}-1\right)\nonumber\\
B_{wind}^{\bar{y}}(t)&=&0\\
B_{wind}^{\bar{z}}(t)&=&0 \nonumber
\end{eqnarray}

Here $B_{wind}^{0}$ is the magnetic field strength in the wind that can be estimated to be $\sim 20\, \mathrm{G}$ \citep{per12}. $k$ is a coefficient defining the steepness of a "step" and $v\approx c$ is the wind propagation speed. We use the same width $d = c \times P_{A}/2 = 3.3 \times 10^{8} \mathrm{cm}$ for the stripes of both polarities since in the equatorial plane one expects both types of stripes to be symmetric.

In our simplified model, the overall magnetic field at time $t$, $B_{tot}(t)$, is a linear addition of a dipole (eq. \ref{dipfield}) and the wind's fields (eq. \ref{windfield}) at that particular moment. In the frame of the dipole:

\begin{equation}
\label{totfield}
B_{tot}(t)=B_{dip} + C \cdot B_{wind}(t)\\
\end{equation}

Here, $C$ is a $3 \times 3$ matrix of the coordinate transformations from the wind to the dipole's frame. Its components can be written as:
\footnotesize
\begin{eqnarray}
\label{transmat}
C_{xx}&=&\sin\phi_{prec}\cos\phi_{spin}-\cos\phi_{prec}\cos\alpha\sin\phi_{spin} \\
C_{xy}&=&-\sin\phi_{prec}\sin\phi_{spin}-\cos\phi_{prec}\cos\alpha\cos\phi_{spin} \nonumber\\
C_{xz}&=&\cos\phi_{prec}\sin\alpha \nonumber\\
C_{yx}&=&\cos\phi_{prec}\cos\theta\cos\phi_{spin}+\sin\phi_{prec}\cos\theta\cos\alpha\sin\phi_{spin} \nonumber \\ &&-\sin\theta\sin\alpha\sin\phi_{spin} \nonumber\\
C_{yy}&=&-\cos\phi_{prec}\cos\theta\sin\phi_{spin}+\sin\phi_{prec}\cos\theta\cos\alpha\cos\phi_{spin}\nonumber\\
&&-\sin\theta\sin\alpha\cos\phi_{spin} \nonumber\\
C_{yz}&=&-\sin\phi_{prec}\cos\theta\sin\alpha-\sin\theta\cos\alpha \nonumber\\
C_{zx}&=&\cos\phi_{prec}\sin\theta\cos\phi_{spin}+\sin\phi_{prec}\sin\theta\cos\alpha\sin\phi_{spin}\nonumber\\
&&+\cos\theta\sin\alpha\sin\phi_{spin} \nonumber\\
C_{zy}&=&-\cos\phi_{prec}\sin\theta\sin\phi_{spin}+\sin\phi_{prec}\sin\theta\cos\alpha\cos\phi_{spin}\nonumber\\
&&+\cos\theta\sin\alpha\cos\phi_{spin} \nonumber\\
C_{zz}&=&-\sin\phi_{prec}\sin\theta\sin\alpha+\cos\theta\cos\alpha \nonumber
\end{eqnarray}
\normalsize
Where $\theta$ is a colatitude of B's spin axis, $\alpha$ is a misalignment of the magnetic axis, and $\phi_{prec}$ and $\phi_{spin}$ are the precessional and spin phases, respectively (see Fig. \ref{fig:pulsargeometry}).

In order to calculate the time-dependent distortions of the field lines, we trace the field lines of the overall magnetic field (eq. \ref{totfield}). This is equivalent to solving the following system of differential equations:

\begin{eqnarray}
\label{diffeqs}
\sqrt{x_{in}^{2}+y_{in}^{2}+z_{in}^{2}}=R_{NS} \nonumber\\
\frac{dx}{B_{tot}^{x}}=\frac{dy}{B_{tot}^{y}}=\frac{dz}{B_{tot}^{z}}
\end{eqnarray}

Here, the first equation represents the initial condition with $R_{NS}$ being the radius of the neutron star.
Due to the complexity of the wind magnetic field (eq. \ref{windfield}), it is impossible to solve the seemingly simple system (\ref{diffeqs}) analytically. Therefore, we use in-house code based on the Runge-Kutta 4 solver to trace the field lines with high precision.

\subsection{Lower Limit of the Emission Height}

\begin{figure}
\centering
\includegraphics*[width=1.0\linewidth]{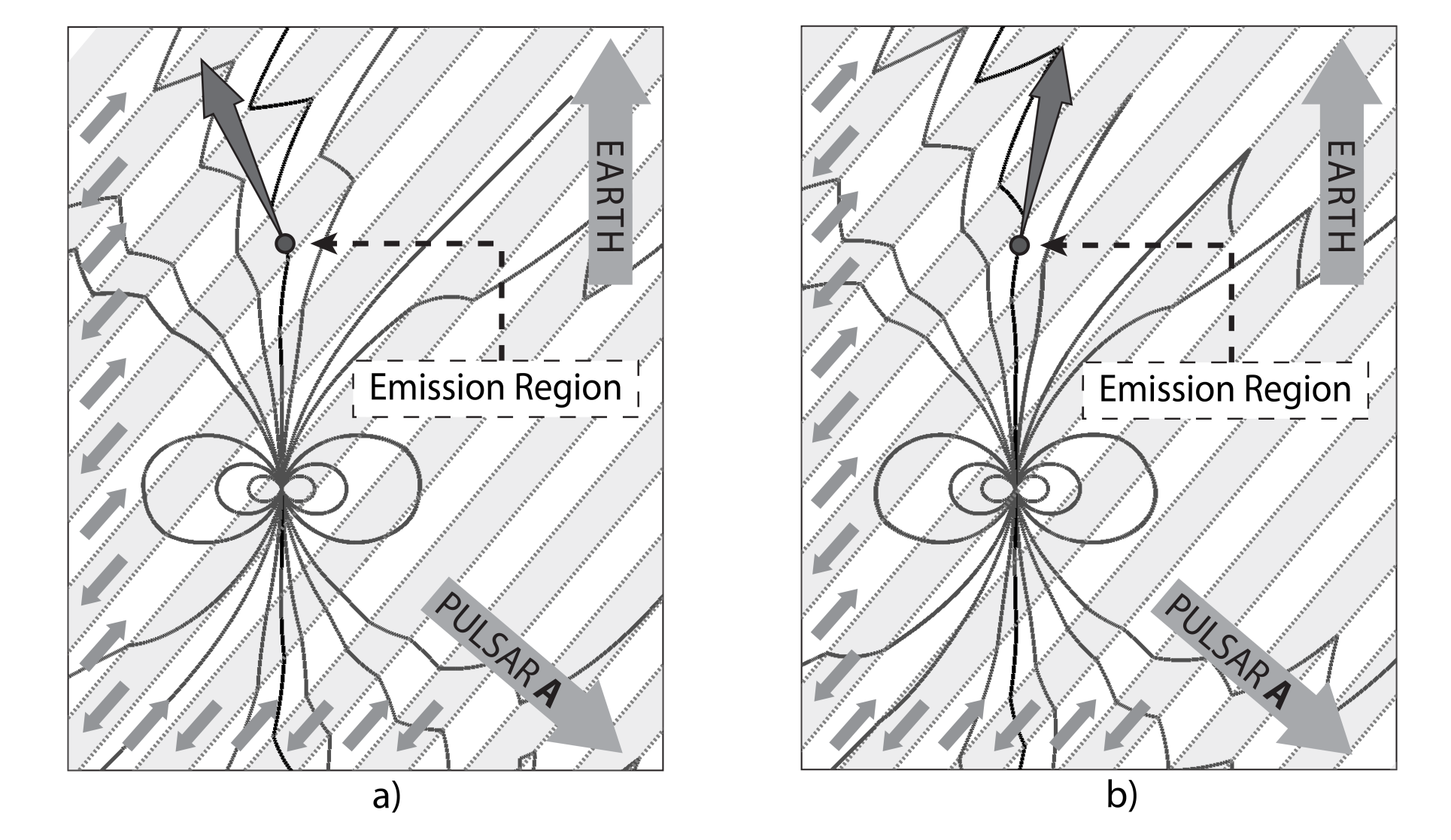}
\caption{Modulation of the emission direction by the striped wind of pulsar A. a) The Dungey-type magnetosphere at some arbitrary time $t$. b) The same magnetosphere after $7\,\mathrm{ms}$. $\beta$ is the angle of incidence.
Shaded stripes represent the regions with the magnetic field of a uniform absolute value but opposite polarity. The small arrows at the borders mark the direction of the magnetic field for each stripe. The strengths of the dipole's and winds magnetic fields were assumed to be equal at $4000 R_{NS}$ along the dipole equator.}
\label{fig:emissiondir}
\end{figure}

Studying the distortions of the polar field line induced by the wind from pulsar A can help us put a lower limit on the radio emission height. Two stripes of the wind with the magnetic fields of opposite polarity cause the deflection of the field line in the opposite directions while passing the same region. This implies that the distortions induced by the passing stripes result in a change of the emission direction (Fig. \ref{fig:emissiondir}), assuming the radio waves are emitted along the tangent to the polar field line. This leads to an observer detecting the periodic intensity fluctuations in the pulse profile (subpulses). The frequency of this modulation will exactly match the spin frequency of pulsar A.

These fluctuations should be large enough to cause a periodic absence of the observed radio emission, which is the reason for the observed subpulse signatures. This can be achieved if the emission region is pushed completely out of the observer's line-of-sight. In other words, the maximum deflection angle at the altitude of the emission generation is larger than the width of the radio emission beam. Thus, subpulses cannot be observed for the values of the emission height for which the deflection amplitude is less than the angular spread of the beam. We use this criteria to estimate the lower limit of the emission height.

In addition to the distance from the star, the amplitude of the deflection angle also depends on $\beta$, the angle of incidence of the wind with respect to the magnetic axis of pulsar B (Fig. \ref{fig:emissiondir}). At the moment of the closest approach (the moment at which the angle between the emission direction and the LOS is minimal), $\beta \approx \phi_{orb} - 180^{\circ}$, where $\phi_{orb}$ is the orbital phase of pulsar B. Therefore, for each orbital phase, there is an altitude below which the deflection is smaller than the value that is necessary for the appearance of the drifting subpulses.

\begin{figure}
\centering
\includegraphics*[width=1.0\linewidth]{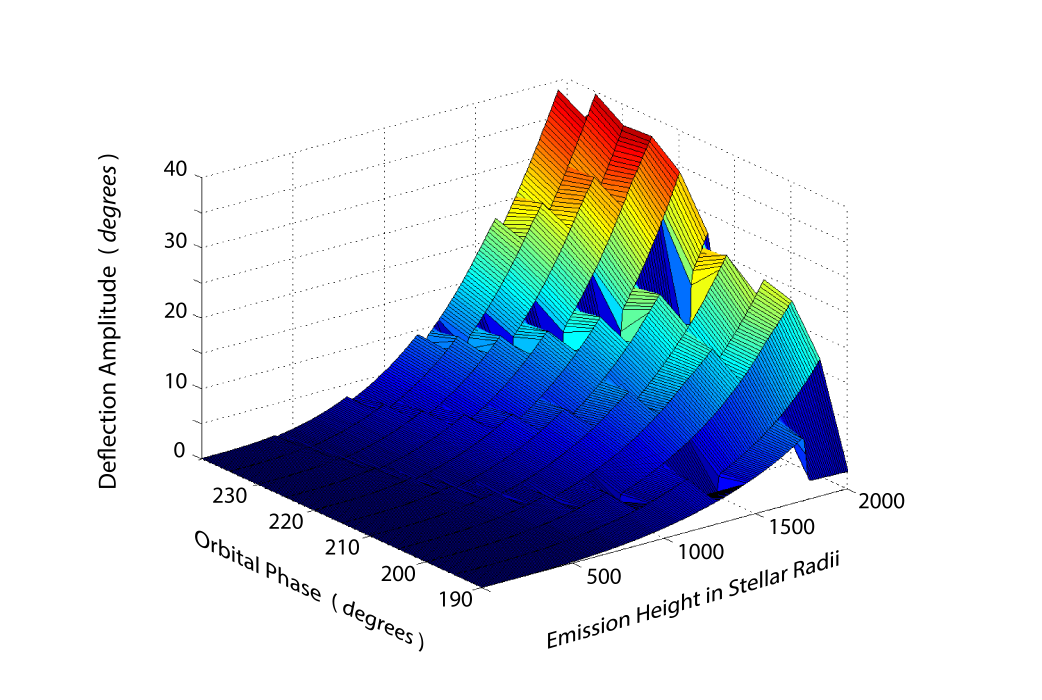}
\caption{Amplitude of the deflection angle for the altitudes up to $2000 R_{NS}$ and orbital phases corresponding to the bright phase 1.}
\label{fig:defl3d}
\end{figure}

The premise of our simulation is to find the minimum distance from the neutron star at which the deflection amplitude is equal to the width of the emission beam. We do this by tracing a polar field line for each orbital phase in BP1 over the period of A. We calculate the maximum angular separation between the local tangents of the polar field line for the different altitudes and orbital phases (Fig. \ref{fig:defl3d}).

Furthermore, by plotting the fixed value contours for the amplitudes of the deflection angle, we can find a minimum allowed emission height for each orbital phase. Additionally, due to the fact that the drifting subpulses were observed only in BP1, we carry out the simulation only for the corresponding orbital phases $\sim [190^{\circ}-240^{\circ}]$.

\begin{figure}
\centering
\includegraphics*[width=1.0\linewidth]{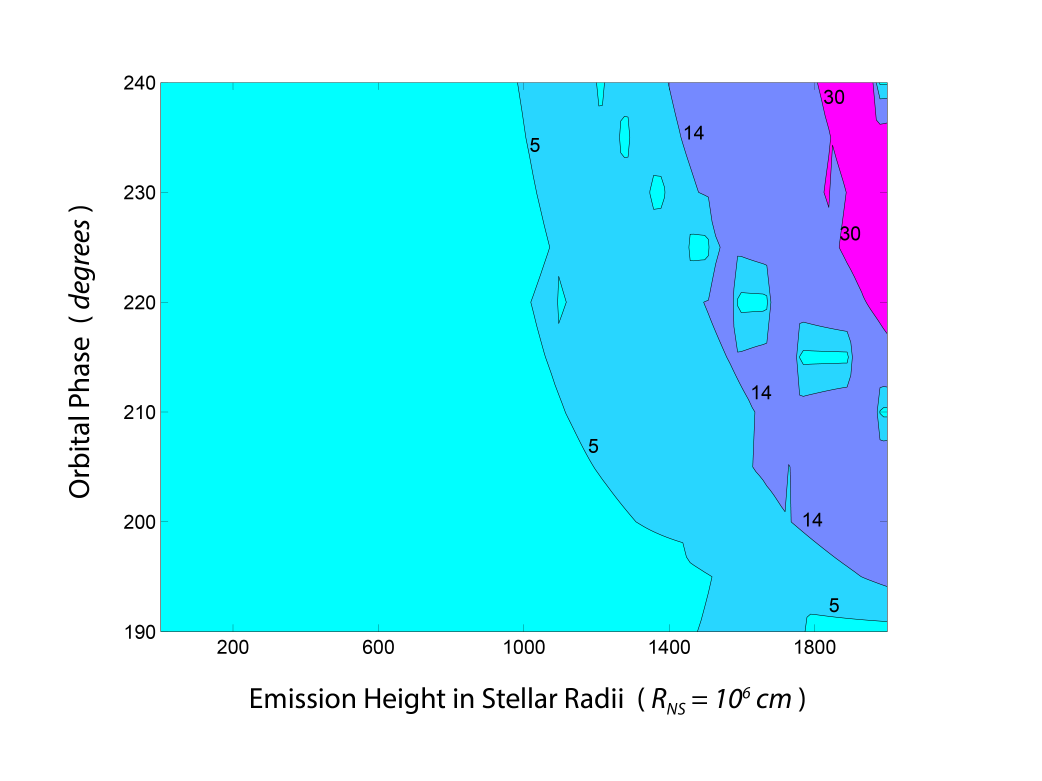}
\caption{Fixed value contours for the amplitude of the deflection angle calculated by using our D61 model for B's magnetosphere. The left-most point of each contour shows the lower limit of the radio emission height and corresponding orbital phase. The $14^{\circ}$ contour indicates the estimated emission height of $\sim 1400 R_{NS}$.}
\label{fig:deflcontours}
\end{figure}

\citet{per10} derived the shape and width of the radio emission beam from the analysis of the long-term evolution of B's pulse profiles. The estimated angular size of the beam, $\sim 14^{\circ}$, corresponds to the overall lower limit of $\sim 1400 R_{NS}$ for the emission height (Fig. \ref{fig:deflcontours}). If we take into account the anisotropy of the beam shape in \citet{per10} and consider only $5^{\circ}$ as a characteristic size of the beam, then we end up with the minimum emission height of $\sim 1000 R_{NS}$ (Fig. \ref{fig:deflcontours}). This emission height is still quite high and can have interesting implications for the models of pulsar radio emission that we discuss in section \ref{sec:Discussion}.

The Dungey-type model of the magnetosphere, discussed in this section, is a simplified way of representing B's overall magnetic field. For instance, one could argue that in a more realistic case when the bow shock is formed due to the wind's impact onto B's magnetosphere, the wind magnetic field should be partially screened and its penetration into the inner magnetosphere should be reduced \citep{mcl04b}. Nevertheless, this model suffices in understanding the occurrence of drifting subpulses in B's radio emission. On the other hand, this method alone cannot explain the absence of the drifting subpulses phenomena in the BP2. However, in the next section we show that when considered together with the MTS confinement model, they are fully consistent with the observational data.

\section{Combined Scenario}

The duration of B's pulse is an order of magnitude larger than the timescale of change of polarity in the incident striped wind, which corresponds to the half period of pulsar A. Therefore, morphology of the orbital bright phases does not depend on the nature of the wind (whether striped or not), but rather depends on the average ram pressure induced by the wind. Thus, our estimate of the radio emission height of $\sim 3750 R_{NS}$ is not expected to change depending on the structure of the wind, since it is inferred from the fitting of the bright phases.

However, a more realistic model of the wind-magnetosphere interaction should account for both phenomena, the jittering of the magnetosphere due to the influence of the striped wind magnetic field (D61 model), and the formation of the paraboloidal boundary encompassing pulsar B's magnetosphere (MTS model). The combination of these two models would allow only part of the striped wind magnetic field to penetrate through the magnetopause of B and reconnect with B's field lines. The penetration parameter $\kappa$, the ratio of the penetrated field to the original field in the wind, ranges from 0 to 1. Its value depends on the properties of the double pulsar system and needs to be explored further in future studies. However, various studies of the solar wind's interaction with the Earth's magnetosphere found the penetration parameter varied over a wide interval (from 0.05 to 0.8) and was strongly dependent on the shape of the magnetopause \citep{kit93, tsyg98b}. Assuming a similar penetration parameter for the double pulsar ($0 \geq \kappa \leq 1 $), the amplitude of the deflection angle for any particular altitude would effectively decrease due to the smaller distorting external magnetic field. As a result, the D61 model estimated emission height of $\sim 1400 R_{NS}$ would increase compared to the system configuration without a screening boundary.

Conversely, we can determine the penetration parameter $\kappa$ from the ratio of the D61 model-estimated and MTS model-estimated emission heights. The latter was determined by using two different approaches: modeling the subpulse drift due to striped wind and fitting the bright phases.

\begin{figure}
\centering
\includegraphics*[width=1.0\linewidth]{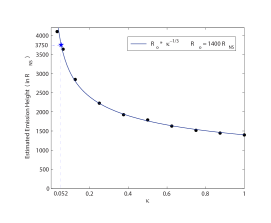}
\caption{Estimated emission height for different values of the penetration parameter $\kappa$. Black circles correspond to the results of our D61 model simulation for different values of the wind magnetic field (equivalent of changing $\kappa$). Solid line represents the best-fit cubic curve. The estimated best-fit value of $\kappa$ corresponding to the emission height of $3750 R_{NS}$ is 0.052.}
\label{fig:kappa}
\end{figure}

As we already mentioned, the magnetopause will screen the part of the wind magnetic field from penetrating into the magnetosphere and distorting the field lines. Thus, instead of the original wind magnetic field  $B_{wind}$, the field lines are distorted by the magnetic field of the reduced strength $\kappa B_{wind}$, where $0 \geq \kappa \leq 1 $ is a penetration parameter. By estimating the emission heights for different values of wind magnetic field, one would essentially find the dependence of emission altitude on $\kappa$. We plotted the results of numerical calculations on the Fig. \ref{fig:kappa}.

We can also find the value of the penetration parameter for which the emission height estimate from the D61 model would be similar to the MTS model estimate. From the Fig. \ref{fig:kappa} we can see that the emission height of $\sim 3750 R_{NS}$ corresponds to $\kappa = 0.052$. Furthermore, the functional dependence of the estimated emission height on the penetration parameter shown on the Fig. \ref{fig:kappa} is in superb agreement with the simple analytic predictions. For the small distortions, the amplitude $\Delta_{defl}$ can be expressed as $\sim \delta B / B$. In our case, if we use the approximate values for $\delta B \approx \kappa B_{wind}$ and $B \approx B_{Dip}(r)$, we get $\Delta_{defl} (r) \sim \kappa B_{wind} / B_{Dip}(r)$. Here, $\kappa =1$ corresponds to the configuration without a screening boundary (considered in the previous section) and $B_{Dip}(r) \sim r^{-3}$. Since our method involves finding $r$ at which $\Delta_{defl} (r)$ equals to the width of the emission beam (which is constant), we arrive to the following expression: $\kappa / r^{3} = const$. This exactly matches the best fit curve on the Fig. \ref{fig:kappa}.

We can use the simple calculations to estimate the magnetic reconnection properties at the boundary of B's magnetosphere. \citet{kit93} found that the coefficient for IMF $B_{y}$'s diffusive penetration into the Earth's magnetosphere can be approximated as $S^{-1/4}$, where $S$ is the Lundquist number. If we use a similar logic, our estimated value of the penetration parameter $\kappa = 0.052$ leads to a Lundquist number of about $\sim 1.3 \times 10^{5}$.

\begin{figure}
\centering
\includegraphics*[width=1.0\linewidth]{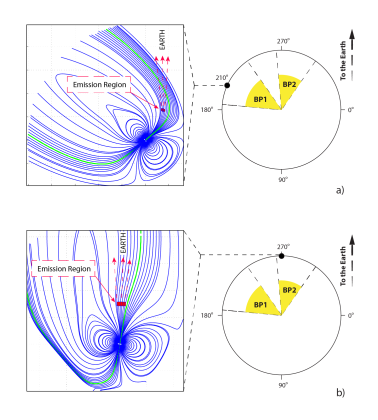}
\caption{Location of the emission region at the moment of the closest approach for two different orbital phases. In both cases, the emission region is located at $3750 R_{NS}$ close to the polar field line. a) In BP1, the radio emission region is closer to the magnetospheric boundary; b) In BP1, emission region moves to the tail of the magnetosphere and therefore further from the boundary.}
\label{fig:ER_location}
\end{figure}

Experimental data also shows the higher penetration parameter in the head part of the Earth's magnetosphere than in the far tail (see Fig. 3 in \citep{kit93}). Owing to the similarities between the the magnetospheres of pulsar B and the Earth, we can assume that the distortions of B's field lines due to the influence of A's wind are stronger in the head part of the magnetosphere (generally closer to the boundary). We can use our MTS model to plot the the position of the emission region with respect to the magnetospheric boundary at the moment of the closest approach for each of the two bright phases. As we can see on the Fig. \ref{fig:ER_location}, the emission region is much further from the magnetopause in BP2 than in BP1 at the moment when the emission direction is the closest to the LOS. This leads to the diminishing distortions in the BP2 due to the weak penetration of external magnetic field for this configuration. Therefore, since the distortions are the primary reason of the B's observed subpulse drift, one expects this phenomena to be absent in the second bright phase, as it has been observed by \citet{mcl04b}.

\section{Discussion} \label{sec:Discussion}

Since the discovery of the double pulsar PSR J0739-3037A/B, a number of approaches have been used to explain the observed orbital modulation of B's radio emission. \citet{jen04} suggested that brightening of B's radio emission is triggered by the $\gamma$-ray emission from A. Their model required a special orientation of A's spin and magnetic axes in order to place the orbital bright phases at their observed longitudes. However, this configuration was inconsistent with the geometry of A, as inferred from the pulse profile evolution \citep{manch05}.
Later, \citet{zhang04} proposed that the emission from B is induced by the energetic particles from the wind of A penetrating the magnetosphere of B. However, as \citet{lyut05bp} pointed out, this mechanism is rendered unsuitable as it does not account for the magnetic bottling effect which reflects the wind particles high above B's magnetosphere.

In this paper, we presented an alternative model of the orbital modulation of pulsar B's radio emission, which is based on the approach proposed by \citep{lyut05bp}. We assumed that B is intrinsically bright at all times but that its emission direction misses our line of sight (LOS) at most orbital phases. However, at certain longitudes, the orbital phase dependent distortions push the emission beam back into our direction, thus causing the effect of brightening. By using this approach, we successfully simulated the bright phases and their evolution at orbital longitudes that closely match the observed longitudes. Additionally, we constructed a more realistic model of B's magnetosphere, and pinpointed the location of the emission region by fitting the morphological properties of the observed orbital modulation (i.e., the bright phases).

Similar to the Sun, pulsar A exerts a powerful wind on its companion B and confines its magnetosphere, forming a cometary tail around it. We constructed a numerical model of B's magnetosphere distortions, which are due to the influence of A's wind. Given the similarities between the double pulsar and the Earth-Sun system, we based our model on the well-accepted semi-empirical model of the Earth's magnetosphere (i.e., T02 model), developed by \citep{tsyg02a,tsyg02b}. The mechanism of B's magnetospheric distortions due to external influence can vary depending on the magnetization of the wind and relative strength of the wind's and pulsar's magnetic fields. Our implementation of the T02 model does not include the effects of the non-zero external wind magnetic field. Rather, B's magnetosphere gets distorted primarily by the wind's ram pressure. Assuming the wind magnetic field has a striped structure, the effects of the distortions due to the reconnection are negated for a timescale larger than the passing time of one stripe (which is essentially the period of A). Therefore, we neglected the reconnection between the wind and B's magnetosphere in the simulations of the bright phases, given that the average pulse duration of $\sim 80\, \mathrm{ms}$ is much larger than the period of A ($\sim 23\, \mathrm{ms}$).

Fitting the bright phases allowed us to estimate the parameters of pulsar B and its magnetosphere. Our obtained value of magnetic inclination $\alpha \approx 56^{\circ}$ is consistent (within error intervals) to the ones inferred by the eclipse modeling \citep{lyut05dp,breton08} and pulse profile evolution analysis \citep{per10}. Furthermore, the estimated absolute value of the spin axis latitude with respect to the orbit, $|90^{\circ} - \theta| \approx 32^{\circ}$, is also consistent with the results of the aforementioned analyses. However, in terms of orientation, our best-fit spin axis direction is nearly antiparallel to the ones estimated by \citet{breton08} and \citet{per10}. This discrepancy might be a result of the degenerate solutions \citet{breton08Th}. However, our model allows an independent estimation of all three axis directions and therefore a unique definition of the angle between the spin axis and the orbital angular momentum (see section \ref{sec:Degeneracies}).

% Very high emission height
\citet{kram08} argued that pulsar B's radio emission should be generated at low altitudes since most of B's magnetosphere is swept away by A's wind. Conversely, our estimated emission height of $0.28 R_{LC}$ is very high. However, even with this high altitude, our model emission region always remains within the boundary of the model magnetosphere ($R \sim 0.9 r_{s}$). Moreover, the emission region is substantially far from the boundary during BP1 and BP2 (see Fig. \ref{fig:ER_location}). Thus, we believe that there is no reason why B's radio emission cannot be generated at higher altitudes. In addition, the orbital modulation of B's radio emission requires orbital phase dependent distortions. However, the distortions are expected to be diminishingly small at lower altitudes. \citet{lyut05dp} and \citet{breton08} inferred from high precision eclipse modeling that the structure of B's magnetosphere is effectively dipolar within $10^{9}\, \mathrm{cm}$. The observational phenomenology of the double pulsar indicates that an emission height above $\sim 1000 R_{NS}$ is more favorable. Moreover, theoretical models of coherent radio emission via beam driven plasma instabilities  place the emission region at very high altitudes $\geq 1000 R_{NS}$ \citep{mach79,lyut98b,lyut99}.

% emission beam structure consistent with theory
We demonstrated that the horseshoe model is a successful fit for the emission beam structure. This model is consistent with the observational data \citep{per10} as well as with the theoretically derived polar cap structure \citep{bai10, wang11}. These force-free theoretical models map the current density distribution in the polar cap region. For the magnetic colatitude of $\alpha = 60^{\circ}$ (which is close to the value for pulsar B), the current density patterns derived by both \citet{bai10} and \citet{wang11} closely matched our best-fit model of the emission beam. This implies that the radio emission is generated on the field lines that carry the maximum current. Therefore, the emission region can be supplied with the relativistic charged particles required for the coherent radio emission generation.

% fixed emission beam shape approx
On the other hand, our implementation of the emission beam implied that the shape and the angular extent of the emission region remain constant. This is a well adopted technique for the isolated pulsars, which do not undergo time-dependent distortions. However, it poses some limitations to the modeling of the double pulsar. The constant emission beam does not allow the model to fully account for the dynamic distortions within B's magnetosphere. This may explain why the simulated relative maximum intensity between the first and second pulse peaks (see Fig. \ref{fig:PPE}) differs from the observed trend \citep{per10}. Namely, the data show that during earlier observations, the leading component of the pulse dominated the trailing component for both bright phases. However, later on, the first peak weakened and eventually fully vanished, but only in BP2.

%disappearance
Given that B's magnetospheric distortions are large due to the high estimated emission altitude, there is an orbital phase at which B's emission beam crosses the LOS for every value of the precession phase. Therefore, simulating the disappearance of B's radio emission was impossible without an imposed cutoff on the emission geometry. We were able to reproduce a gradual decay of B's intensity over the observed time-range by limiting the possible emission generation sites to the vicinity of the null-charge surface. This result may be useful when probing the radio emission mechanisms and the detailed structure of the pulsar magnetosphere in future research. However, more precise fitting of the cutoff parameters would be required.

% Subpulse drift
In addition to the confinement model, we constructed a simple reconnection model for the wind-magnetosphere interactions. Based on the Earth's magnetosphere model by \citep{dungey61}, we represented the magnetosphere of B by a simple superposition of a point dipole and the striped wind from A. The addition of the striped wind resulted in the periodic distortions of B's dipole, which we quantified by the deflection angle of the field lines. This allowed us to successfully reproduce the drifting subpulse features in B's radio emission, and thus to confirm that this phenomenon is indeed induced by A's influence. The assumption of the striped structure of the magnetic field in A's wind is crucial to the reproduction of the subpulse drift features at the right periodicity.

Additionally, from the comparison of the results of the D61 and MTS models, we constrained the reconnection properties at the wind-magnetosphere interaction boundary. We estimated the magnetic field penetration coefficient $\kappa$ as $\sim 5\%$. It should be noted that satellite and ground-based observations reported similar values of $\kappa$ for the Earth \citep{kit93,tsyg07}.

Partial screening of the wind magnetic field by the boundary of B's magnetosphere combined with the high emission altitude provided a feasible explanation of the absence of drifting subpulses in BP2. Given the high emission height, the observed emission is generated much closer to the boundary for BP1 than for BP2 (see Fig. \ref{fig:ER_location}). Therefore, an observer would detect more prominent striped wind signatures (i.e., subpulse drift) during BP1 than during BP2. This exact scenario has been observed by \citet{mcl04b}, indicating that A's striped wind influences B's magnetosphere and that radio emission is generated at high altitudes.

% inconsistency with GR?

% simplifications
We used a number of simplifications in our modeling of B's magnetosphere. Even though our model included the contributions from all Earth's known magnetospheric currents, it did not consider the effects of pulsar rotation. The distortions due to the relativistic electromagnetic effects (such as field line sweepback, abberation, and retardation) become significant at higher altitudes \citep{shitov85,bltz91,bai10}. However, we tried to compensate for this fact by imposing a retardation effect with arbitrary polynomial scaling on the field lines. The obtained best-fit retardation angle increases quadratically with distance from the pulsar ($\sim 2.2 \left(r/R_{LC}\right)^{2}$). However, it is highly probable that we could improve the fit of the bright phases, as well as the pulse profile evolution, by considering a more general form of the deflections \citep{bai10}.

% Reappearance time

Pulsar B is expected to reappear sometime within its precession period of $\sim 75$ years, assuming that the primary reason of its disappearance was the spin axis precession. The exact reappearance date, however, varies depending on the model of B's magnetosphere and the emission beam. For a point dipole magnetosphere with an elliptical horseshoe-shaped emission beam, the predicted reappearance time is around the year 2035 \citep{per10}. The same analysis, but with the two-sided horseshoe beam, yields the estimated date in the year 2014. \citet{breton08Th} used a similar setup with a $20^{\circ}$-wide beam to predict the reappearance time of B. He concluded that the reappearance will happen in the year 2024 for the emission from the same pole and in the year 2033 for the emission from the opposite pole.

Since our model of B's magnetosphere is much more complex than the point dipole, our predictions of B's reappearance date differ from the ones by \citet{per10} and \citet{breton08Th}.
We predict three possible epoches for B's reappearance. For the two-pole emission configuration, we expect pulsar B to become observable again in the year 2043, if the radio emission is generated near the null-charge surface only. However, if the emission is generated in the whole region where $\Omega \cdot \mathrm{B} \geq 0$, then B is expected to reappear in 2034. For a single-pole emission configuration, both cutoffs yield the same reappearance date of 2066. Therefore, the reappearance date of B is indicative of which of these pulsar magnetospheric properties are at play. We will be able to find out if both poles produce emission beams by the year 2034, and if the emission is only generated near the null-charge surface before the year 2066.

\section{Summary} \label{sec:Summary}

We assumed that PSR J0737-3039B was intrinsically bright along the whole orbit and that all of its observed modulation patterns are due to the phase dependent deflections of the radio emission direction. These deflections are a result of the influence of A's wind on B's magnetosphere, which resembles the solar wind's influence on the Earth's magnetosphere. We used this similarity between the double pulsar and the Earth-Sun system to develop the model for B's distorted magnetosphere. We used two contrasting but complementary models of the Earth's magnetosphere to reproduce the observed features of pulsar B's radio emission. Namely, we reproduced the time-evolving orbital bright phases with up to a $1^{\circ}$ precision by using our Modified Tsyganenko's model (MTS), based on the semi-empirical confinement model by \citet{tsyg02a,tsyg02b}. Additionally, we were able to explain the subpulse drift phenomena observed in the first bright phase by using the reconnection model of planetary magnetospheres developed by \citet{dungey61}.

In addition, the analysis of the bright phases along with the secular evolution of B's pulse profile \citep{per10} resulted in a partial (horseshoe shaped) elliptical emission beam with a half opening angle of $13^{\circ}$ (along the semi-major axis of the ellipse). The corresponding footpoints of the beam field lines are arranged in a similar shape on the side of the polar cap region that is directly across from the spin axis, with respect to the magnetic pole. The obtained shape and position of the horseshoe closely mimic the current distribution pattern in the polar cap, as was inferred from theoretical models of pulsar magnetospheres \citep{bai10,wang11}. Therefore, we concluded that radio emission follows the same pattern as the current density distribution in B's magnetosphere.

We also estimated the radio emission height in pulsar B. The orbital phase dependent distortions of B's magnetosphere, which are responsible for the occurrence of the bright phases, are very sensitive to the altitude of the emission region. Therefore, by fitting the observed morphology of the bright phases, we were able to pinpoint  the location of the emission generation region. Given that emission height is one of the main probes of coherent radio emission mechanisms \citep{mel95}, our result of $R \sim 0.3 R_{LC}$ makes the polar cap emission models unsuitable, at least for pulsar B. However, the estimated location of the emission region is consistent with some of the beam instability driven models of the coherent radio emission (e.g. \citep{mach79,lyut98b,lyut99}.

Alternatively, we estimated the radio emission altitude by studying B's drifting subpulses. The observed properties of this phenomenon imply that there is a strong influence of A's wind on B's emission region, at least in bright phase 1. We suggested that the modulation of B's radio emission at the period of A is due to the reconnection of B's field lines with the magnetic field in A's wind. The field line distortions due to the reconnection push the direction of B's radio emission in and out of our line of sight, in turn, this results in the observed intensity variations. We were able to constrain the emission height in pulsar B by using the reconnection model of the wind-magnetosphere interaction developed by \citet{dungey61}, given that the amplitude of these deflections depends on the ratio of the local strengths of the two fields, and hence on the distance from the pulsar.

Moreover, our analysis of the combination of the two models (the Dungey-type reconnection model and the confinement model with a paraboloidal boundary) led to the derivation of another interesting property of the wind-magnetosphere interaction in the double pulsar. The combination of the results of the two models indicates that only $5\%$ of the wind magnetic field penetrates the dayside part of B's magnetopause. The penetration coefficient further decreases towards the tail of the magnetosphere. This can explain the absence of the subpulse drift in the second bright phase, given that in BP2 the observed emission is generated in the tail of the magnetosphere and therefore is less affected by the magnetic field in A's wind.

Furthermore, we resolved all but one degeneracy in pulsar B's geometry. Even though the inclination of the orbit is precisely inferred from the measured Shapiro delay \citep{kram06}, the orientation of the orbital angular momentum with respect to the plane of sky remains undetermined. However, we were able to determine the direction of the spin axis and the geodetic precession with respect to the orbital angular momentum. Further improvements to this result can be achieved by refining the resolution of the simulation grid.

We successfully simulated the disappearance of B's radio emission by implementing the spatial cutoff of the emission generation within a pulsar magnetosphere. By fitting the observed rate of disappearance, we concluded that B's detected radio emission was generated in the vicinity of its null-charge surface.

We predicted different reappearance dates depending on whether one or both poles contribute to the pulsar radio emission. In addition, the predicted reappearance date changed for different models of the cutoff. For a two-pole emission configuration, pulsar B is expected to reappear either in the year 2034 or year 2043 depending on the two different models of the cutoff (see section \ref{sec:Reappearance}). On the other hand, for a single-pole configuration, both models of the cutoff yield the same reappearance date in the year 2066, albeit with a different rate. We will be able to determine which of these arrangements is realized in pulsar B by examining the date and rate of B's reappearance, once it becomes observable again.

\section*{Acknowledgements}
We thank Kostas Gourgouliatos, Benetge Perera,
and Maura McLaughlin for very useful discussions. We also thank Nikolai Tsyganenko and Rene Breton for their help and advice. This work was supported in part by NASA grant NNX09AH37G.

\begin{comment}
\end{comment}

\end{document}